\documentclass[twocolumn,floats,showpacs,superscriptaddress,pre]{revtex4}
\usepackage{amssymb}
\usepackage{graphicx}
\usepackage{dcolumn}
\usepackage{amsmath}
\usepackage{bm}
\usepackage{epsfig}

\def\defi{{\buildrel \;def\; \over =}}
\newcommand{\be}{\begin{equation}}
\newcommand{\ee}{\end{equation}}

\newcommand{\mediaT}[1]{\left\langle #1 \right\rangle}

\newcommand{\media}[1]{\langle #1 \rangle}

\begin{document}

\title{Communication and correlation among communities}
\author{M. Ostilli }
\affiliation{Departamento de F{\'\i}sica da Universidade de Aveiro, 3810-193 Aveiro,
Portugal}
\affiliation{Center for Statistical Mechanics and Complexity, 
INFM-CNR SMC, Unit\`a di Roma 1, Roma, 00185, Italy.}
\author{J. F. F. Mendes}
\affiliation{Departamento de F{\'\i}sica da Universidade de Aveiro, 3810-193 Aveiro,
Portugal}

\begin{abstract}
Given a network and a partition in communities, 
we consider the issues ``how communities influence each other''
and ``when two given communities do communicate''. 
Specifically, we address these questions in the context of small-world networks,
where an arbitrary quenched graph is given and long range connections are randomly added. 
%To this aim we consider a recently proposed method to analyze models defined on small-world graphs \cite{SW}
%to include $n$ interacting small-world communities with arbitrary sizes and disorders. 
We prove that, among the communities, 
a superposition principle applies and gives rise to a natural generalization
of the effective field theory already presented in [Phys. Rev. E 78, 031102] ($n=1$),
which here ($n>1$) consists in a sort of effective TAP (Thouless, Anderson and Palmer) equations 
in which each community plays the role of a microscopic spin. 
The relative susceptibilities derived from these equations calculated at finite or zero
temperature, where the method provides an effective percolation theory, give us 
the answers to the above issues.
%As for $n=1$, this generalization  
%is exact in the paramagnetic regions (at $T=0$ this means below the percolation threshold) and provides 
%effective approximations in the other regions. 
Unlike the case $n=1$,
asymmetries among the communities may lead, via the TAP-like structure of the equations,    
to many metastable states whose number, 
in the case of negative short-cuts among the communities,
may grow exponentially fast with $n$.
As examples we consider the $n$ Viana-Bray communities model 
%(which includes the $n$-Curie-Weiss communities and  
%the $n$-Sherrington-Kirkpatrick communities models as special limits),
and the $n$ one-dimensional small-world communities model.
Despite being the simplest ones, the relevance of these models in 
network theory, as \textit{e.g.} in social networks, is crucial and no
analytic solution were known until now. Connections between percolation and the fractal dimension of a network are also discussed. 
Finally, as an inverse problem, we show how, from the relative susceptibilities,
a natural and efficient method to detect the community structure of a generic network arises.

For a short presentation of the main result see arXiv:0812.0608.    
\end{abstract}

%\pacs{05.50.+q, 87.18.Sn, 64.70.-p, 64.70.P-}
\maketitle

\email{ostilli@roma1.infn.it}

\section{Introduction} \label{intro}
In the last decade we have seen an impressive growth of the network's science
and of its broad range of applications in fields as diverse as physics, 
biology, economy, sociology, neuroscience, etc...
\cite{DM1,DM,Vespignani,NewBar,Guido}. 
Many analytical and numerical methods to investigate the statistical properties 
of networks, such as degree distribution, clustering coefficient, percolation, 
and critical phenomena at finite temperature, as well as dynamical processes, are nowadays available
(see \cite{Review} and references therein). 
In particular, in recent times, the issue to find
the ''optimal'' community's structure that should be present in a given random graph (a network)
$(\mathcal{L},\Gamma)$, $\mathcal{L}$ and $\Gamma$ being the set of the vertices
and of the bonds, respectively, has received much attention. 
The general idea behind the community's structure of a given network comes from the observation
that in many situations real data shows an intrinsic partition of the vertices of the graph in $n$ groups, called
communities, $\mathcal{L}=\cup_{l=1}^{n}\mathcal{L}^{(l)}$, such that between any two communities
there is a number of bonds that is relatively small if compared with the number of bonds present in each community. 
If we indicate by $\Gamma^{(l,k)}$ the set ob bonds connecting the $l$-th and the $k$-th
communities, we can formally express the above idea 
by using the decomposition $\Gamma=\cup_{l\leq k=1}^{n}\Gamma^{(l,k)}$, 
and the inequality $|\Gamma^{(l,k)}|\ll |\Gamma^{(l)}|,|\Gamma^{(k)}|$, for $l\neq k$.
The partition(s) can be used to build a higher-level meta-network where the meta-nodes are now the communities  
(cells, proteins, groups of people, $\ldots$) and play
important roles in unveiling the functional organization inside the network.
In order to detect the community's structure of a given network, many methods have been proposed 
and special progresses have been made by mapping the problem for identifying community structures to
optimization problems 
\cite{Blatt,Girvan,Newmanfast,Clauset,Caldarelli,Guimera,Jorg}, by looking for $k-$clique sub-graphs \cite{Palla},
or by looking for clustering desynchronization \cite{Boccaletti} and, very recently, by using
random walks \cite{Lambiotte}.
In general there is not a unique criterion to find the community's structure \cite{Gulbahce}. However, once obtained
some structure, whatsoever the method used, 
and assuming that the found partition $(\cup_{l=1}^{n}\mathcal{L}^{(l)},\cup_{l\leq k=1}^{n}\Gamma^{(l,k)})$ 
represents sufficiently well the intrinsic community's structure of the given network \cite{Newmanrob}, 
there is still left
the fundamental issue about the true relationships among these communities.
Under which conditions, 
and how much two given communities communicate, how they influence each other, positively or negatively,
what is the typical state of a single community, what is the expected behavior for $n$ large, etc... are all issues
that cannot be addressed by simply using the above methods to detect the community structure.
In fact, all these methods, with the exception of Refs. \cite{Blatt}, \cite{Boccaletti}, and \cite{Lambiotte}, 
are essentially based only on some topological analysis 
of the network, and in most cases, only local topological properties are taken into account. 
The way to uncover the real communication among the communities
is to pose over the graph $(\mathcal{L},\Gamma)$ a minimal model in which the vertices
assume at least two states, \textit{i.e.}, as the spins in an Ising model. 
Confining the problem to the equilibrium case 
we have hence to use the Gibbs-Boltzmann statistical mechanics and find the relative susceptibilities $\chi^{(l,k)}$
among the communities of a suitable Ising model. 
In this approach the temperature $T$ can be seen
as a parameter describing the freedom of the vertices to assume a state independently of the state of the other vertices,
while the coupling $J_{i,j}^{(l,k)}$ between two vertices $i$ and $j$ belonging to the $l$-th and $k$-th
community, respectively, as a tendency of the vertices to be positively or negatively correlated,
according to the amplitude and to the sign of $J_{i,j}^{(l,k)}$.

We point out that, given a community structure, our main aim is to calculate
the magnetizations $m^{(l)}$ and the relative
susceptibilities $\chi^{(l,k)}$ of the communities, while 
Refs. \cite{Blatt}, \cite{Boccaletti} and \cite{Lambiotte}, 
treat the quite different problem of detecting the community structure by looking for the partition of the graph
that, among the communities, minimizes the correlations, the synchronization, or the diffusion, respectively. 
Although this is a natural and interesting way for defining a community structure, and to which
we devote a study in this paper too and find some connections with \cite{Lambiotte}, in many situations
the obtained partition does not correspond to the intrinsic partition of the graph~\footnote{
Concerning for example \cite{Blatt}, where one studies the correlation function of an unfrustrated
$q$-states Potts model, requiring that the partition 
$(\cup_{l=1}^{n}\mathcal{L}^{(l)},\cup_{l\leq k=1}^{n}\Gamma^{(l,k)})$ found in this way coincides with the
intrinsic community structure of the given graph, would imply that the vertices of the graph
could be truly described as $q$ states variables, while in general, vertices can have much more complex
functions in the graph, and a description in terms of $q$ states can be effectively used to represent
only some of their functions as, \textit{e.g.}, the function of communication.}.

At least in principle, if a Gibbs-Boltzmann $\exp(-\beta H)$ distribution with some Hamiltonian $H$ has
been assumed, one can obtain $\beta J_{i,j}^{(l,k)}$ from the data of the given graph by isolating
the two vertices $i,j$ from all vertices of the graph other then them, and by measuring the correlation function
of the obtained isolated dimer $\mediaT{\sigma_i\sigma_j}'$, where $\mediaT{\cdot}'$ stands for the Gibbs-Boltzmann
average of the isolated dimer 
~\footnote{
However, concerning the communication properties, 
as will become clear later, if one knows that all the couplings associated to $\Gamma$ are in average strictly positive, 
at zero temperature the theory provides a general result which is completely independent of the couplings.}.
The general problem is actually more complicated due to the presence of two sources of disorder since
both the set of the bonds $\Gamma$, and the single couplings $\{J_{i,j}^{(l,k)}\}$, may change with time.
Assuming that the time scale over which these changes take place is much larger than that 
of the thermal vibrations of the spins, we have then to facing a disordered Ising model 
with quenched disorder. 

In this paper we specialize this general problem to the case of Poissonian disorder of the
graph, while we leave the disorder of the couplings arbitrary. We formulate the problem in
terms of Ising models on generic small-world graphs \cite{Watts}: given an arbitrary graph $(\mathcal{L}_0,\Gamma_0)$,
\textit{the pure graph}, and
a community's structure $(\cup_{l=1}^{n}\mathcal{L}_0^{(l)},\cup_{l\leq k=1}^{n}\Gamma_0^{(l,k)})$,
in which each community has an arbitrary size, we consider a generic Ising Hamiltonian $H_0$ defined 
on this non random (quenched) structure, \textit{the pure model}, characterized by arbitrary couplings $J_0^{(l,k)}$,
and we add some random connections (short-cuts) 
with average connectivities $c^{(l,k)}$, along which a random coupling $J^{(l,k)}$ takes place,
and study the corresponding random Ising model, \textit{the random model}, having therefore a random Hamiltonian $H$.

In \cite{SW} we established a new general method to analyze critical phenomena on small-world models:
we found an effective field theory that generalizes the Curie-Weiss mean-field theory
via the equation 
\begin{eqnarray}
\label{THEO0}
m^{(\Sigma)}=m_0(\beta J_0^{(\Sigma)},\beta J^{(\Sigma)}m^{(\Sigma)}), 
\end{eqnarray}
and that is able to take into account both
the infinite and finite dimensionality simultaneously present in small-world models.
In Eq. (\ref{THEO0}), $m_0(\beta J_0;\beta h)$ represents the
magnetization of the pure model, \textit{i.e.}, without short-cuts, 
supposed known as a function of the short-range coupling $J_0$ and arbitrary external field $h$,
whereas the symbol $\Sigma$ stands for the ferro-like solution, $\Sigma=F$, or the spin glass-like solution, $\Sigma=SG$,
and $J_0^{(\Sigma)}$ and $J^{(\Sigma)}$ are effective couplings.
Here we generalize this result to the present case of $n$ communities of arbitrary sizes and interactions; 
short-range and long-range (or short-cuts) couplings.
We show that, among the communities, a natural superposition principle applies and
we find that the $n$ order parameters, F or SG like, obey a system  
of equations which, a part from the absence of the Onsager's reaction term \cite{Onsager},
can be seen as an $n\times n$ effective system of TAP (Thouless, Anderson and Palmer) equations \cite{TAP} in which each community
plays the role of a single ``microscopic''-spin $m^{(\Sigma;l)}$, $l=1,\ldots,n$ and,
depending on the sign of the couplings, behave as spins immersed in a ferro or glassy material.
%through its own order parameter, $m^{(\Sigma;l)}$, $l=1,\ldots,n$.

As for one single community, our method
is exact in the paramagnetic region (P) 
(more precisely is exact in the region where any order parameter is zero) 
and provides an effective approximation in the other regions,
becoming exact for unfrustrated disorders even off the P region in the
limits $c^{(l,k)}\to 0^+$ and $c^{(l,k)}\to \infty$.
In the Ref. \cite{SW} ($n=1$) we established the general scenario of the critical behavior
coming from these equations, stressing the differences between the cases $J_0\geq 0$ and $J_0<0$,
the former being able to give only second-order phase transitions with classical critical exponents, 
whereas the latter being able to give rise, for a sufficiently large connectivity $c$, to 
multicritical points with also first-order phase transitions.  
The same scenario essentially takes place also for $n\geq 2$ provided that the $J_0$'s and the $J$'s be 
almost the same for all the communities (and in fact in this case, \textit{i.e.}, near the homogeneous case, 
the partition in $n$ communities does not turn out to be very meaningful and taking $n=1$ would lead to 
almost the same result), 
otherwise many other situations are possible. In particular, unlike the case $n=1$, relative antiferromagnetism
between two communities $l$ and $k$ is possible as soon as the $J^{(l,k)}$ have negative averages,
while internal antiferromagnetism inside a single community, say the $l$-th one, due to the presence of
negative couplings $J_0^{(l)}<0$, is never possible as soon as disorder is present. 
Less intuitive and quite interestingly, if we try to connect randomly with some added connectivity $c^{(l,k)}$
the $l$-th community having inside only positive couplings (``good'') to the $k$-th community having
inside only negative couplings (``bad''), not only the bad community gains
a non zero order, but even the already good community gets an improved order. 

However, with respect to the case $n=1$, another peculiar feature to take
into account is the presence of many metastable states. In fact, this is a general 
feature of the TAP-like structure of the equations: as we consider systems with an increasing number of communities,
the number of metastable states grows with $n$ and may grow exponentially fast in the case
of negative short-cuts. A metastable state can be made virtually stable (or, more precisely leading)
by forcing the system with appropriate initial conditions, by fast cooling, or by means of suitable external fields.
As a result, with respect to variations of the several parameters of the model 
(couplings, connectivities, sizes of the communities), 
the presence of metastable states may lead itself to first-order phase transitions
even when the $J_0$ are all non negative. 
This general mechanism has been already
studied in the simplest version of these models, namely the $n=2$ Curie-Weiss model 
($J_0=0$ and $c^{(1,2)}=\infty$),
where a first-order phase transition was observed to be tuned by the relative
sizes of the two communities and by the external fields \cite{Contucci}; 
moreover, first-order phase transitions have been observed
in numerical simulations of a two dimensional small-world model with directed shortcuts \cite{Sanchez}~\footnote{
Even if, for simplicity, we formulate our small-world models only through undirected shortcuts,
as we shall show, 
the effective long-range couplings $J$'s entering the self-consistent equations and
connecting two different communities are directed couplings (that is non symmetric)
when the sizes of the two communities differ.}.
In particular, in system of many communities,
$n\gg 1$, a remarkable and natural presence of first-order phase transitions (tuned by the several parameters) 
is expected which, if the $J$'s or the $J_0$'s are negative, 
reflects on the fact that the communities, at sufficiently low temperature, behave
as spins in an effective glassy state \cite{Parisi,Fisher}.

Finally, we show that the theory can be projected at zero temperature where a natural
effective percolation theory arises. Then, in this limit, a quantity of remarkable importance,
the relative susceptibility among the communities, is provided and we will show that, by starting from the data of the network, 
it can be efficiently sampled via simulated annealing procedures.
Such a quantity in fact tells us in a not ambiguous manner whether two given communities $l$ and $k$ do communicate or not, and what is their
characteristic time $t^{(l,k)}$ to exchange a unit of information. It will result clear that, given the pure graph, unlike a local analysis 
(based therefore only an elementary use of the adjacency matrix) might say,  
the presence of some bonds $\Gamma_0^{(l,k)}$ between the two communities, does not guarantees that 
they communicate, \textit{i.e.}, that be $t_0^{(l,k)}<\infty$. 
More in general, it will become clear that even a minimal model such as the one we introduce, due to the fact that 
it incorporates exactly all the correlations, short- and long-range like, can give rise to situations
which drastically differ from methods in which only a local analysis of the bonds is taken into account
and/or correlations (including their signs) are never introduced.

As mentioned before, as a byproduct,
we show also that, in particular, the percolation theory 
provides itself another natural way to detect community's structures.
More precisely, similarly to what done in \cite{Lambiotte}, we can define
a family of generalized modularity functions \cite{Girvan} able to probe
the community structure of the given network, pure or random, at several
length scales. We will see that the algorithm of this method turns out to be 
statistically efficient in the limit of small and infinite length scales.

In this paper, as first analytical applications of the method,
we consider two important class of models: the generalized Viana-Bray (VB) model \cite{Viana} and its
special limits of infinite connectivity, \textit{i.e.}, the generalized Curie-Weiss (CW) and
the generalized Sherrington-Kirkpatrick (SK) models \cite{SK}; and the generalized one-dimensional small-world 
models for $n$ communities. A complete analysis of these two class
of models is beyond the aim of this paper since a deeper study, 
also equipped with some numerical analysis
of the self-consistent equations and, more in general, of the minima of the associated
Landau free energy density, would be required. 
We point out however that our results are completely novel. 
Notice in fact that, without any intention to be exhaustive  in citing 
the large literature on the subject,
the state of the art of analytical methods for disordered Ising models 
defined over Poissonian small-world graphs results nowadays as follows: 
\textit{i)} in the case of no short-range couplings, $J_0=0$, and for one community, $n=1$,
modulo a large use of some population dynamics algorithm for low temperatures,
the replica method and the cavity method \cite{Parisi,MezardP,Diluted,Masayuki2} have established the base
to solve exactly the model in any region of the phase diagram,
even rigorously in the SK case \cite{Guerra,Talagrand} and in unfrustrated cases \cite{Luca};
\textit{ii)} for $J_0\neq 0$ and $n=1$ these methods have been successfully applied to the
one-dimensional case \cite{Niko,Niko2} but a generalization to higher dimensions 
(except infinite dimensions \cite{Masayuki}) seems impossible
due to the presence of loops of any length~\footnote{Though perturbative loops expansions around the
tree like approximation are possible, see \cite{Montanari,PS,CC}.}; on the other hand, 
even if it is exact only in the P region, the method we have presented 
in the Ref. \cite{SW}, modulo solving analytically or numerically a non random Ising model, 
can be exactly applied in any dimension,
and more in general to any underlying pure graph $(\mathcal{L}_0,\Gamma_0)$;
\textit{iii)} for $J_0=0$ and $n\geq 2$, the problem was solved
only in the limit of infinite connectivity: exactly in the $n=2$ CW case in its general form, 
which includes arbitrary sizes of the two communities, but with no coupling disorder \cite{Contucci}; 
and, within the replica-symmetric solution, in the generic $n$ SK case, but only in the presence 
of a same mutual interaction among the $n$ communities of same size \cite{Almeida,MOI}.
Out of this range of models, no method was known to face
analytically the general problem with finite connectivities, in arbitrary dimension $d_0$, and with a general disorder, 
despite its relevance in network theory, as \textit{e.g.}, in social networks~\footnote{See
comments and references reported in \cite{Contucci}.}.

The paper is organized as follows.
In Sec. II we introduce the small-world communities network over which
we define the random Ising models.
In Sec. III we present the result: in Sec. IIIA we provide the self-consistent equations,
the correlation functions, the Landau free energy density and the relative susceptibilities;
in Sec. IIIB we analyze the phase transition scenario;
in Sec. IIIC we discuss the level of accuracy of the method.
In Sec. IV we apply the method to the above mentioned example cases (CW, SK, VB and one dimensional models).
In Sec. V we consider the theory at zero temperature obtaining the percolation theory
and the characteristic times of communication among communities. In this section, as byproducts, we show an interesting
connection with the concept of fractal dimension and how to detect a community structure within our framework. 
Secs. VI an VII are devoted to the proof. 
Finally in Sec. VIII some conclusions are drawn.

A short presentation of this work can be found in arXiv:0812.0608.
%
%%%%%%%%%%%%%%%%%%%%%%%%%%%%%%%%%%%%%%%%%%%%%%
%

\section{Random Ising models on small-world communities}
\label{models}
Let be given $n$ distinct graphs $(\mathcal{L}_0^{(l)},\Gamma_0^{(l)})$, $l=1,\ldots,n$,
$\mathcal{L}_0^{(l)}$ and $\Gamma_0^{(l)}$ being the set of vertices and bonds
of the $l$-th graph, respectively~\footnote{In the Ref. \cite{SW} we were mainly interested
in the cases in which $(\mathcal{L}_0^{(l)},\Gamma_0^{(l)})$ is a regular lattice of dimension $d_0$, 
we recall, however, that there is no restriction in the choice of the graph.}.
Elements of a set of vertices $\mathcal{L}_0^{(l)}$ will be indicated with Latin index $i$ or $j$, whereas elements 
of a set of bonds $\Gamma_0^{(l)}$ will be indicated as couples $(i,j)$.
Let the size of $\mathcal{L}_0^{(l)}$ be 
\begin{eqnarray}
|\mathcal{L}_0^{(l)}|=N^{(l)}=\alpha^{(l)}N,
\label{alpha}
\end{eqnarray}
where the $\alpha^{(l)}$'s are $n$ non negative numbers such that~\footnote{For shortness, from now
on, all the sums over the graph index will be understood to run from $1$ to $n$.}
\begin{eqnarray}
\sum_l\alpha^{(l)}=1.
\label{alpha1}
\end{eqnarray}
Moreover, let be given other $n(n-1)/2$ distinct graphs $(\mathcal{L}_0^{(l,k)},\Gamma_0^{(l,k)})$, $l<k$, 
with $l,k=1,\ldots,n$,
where $\mathcal{L}_0^{(l,k)}\defi \mathcal{L}_0^{(l)}\cup \mathcal{L}_0^{(k)}$ is the sum-set of the vertices
of $\mathcal{L}_0^{(l)}$ and $\mathcal{L}_0^{(k)}$, 
and $\Gamma_0^{(l,k)}$ an arbitrary set of bonds connecting some vertices of $\mathcal{L}_0^{(l)}$
with some vertices of $\mathcal{L}_0^{(k)}$.

Given, for each community, an Ising model - shortly \textit{the pure model} of the community
with Hamiltonian 
\begin{eqnarray}
H_0^{(l)}\defi -\sum_{(i,j)\in \Gamma_0^{(l)}}J_{0;(i,j)}^{(l)}\sigma_{i}\sigma_{j}-h^{(l)}
\sum_{i\in\mathcal{L}_0^{(l)}} \sigma_i, %\quad l=1,\ldots,n, 
\label{H0l}
\end{eqnarray}
let be
%~\footnote{We will be mainly interested in the case $J_{0;(i,j)}^{(l,k)}=0$
%for $l\neq k$, however, for not loosing generality we consider the possibility
%that there are also short-range interactions between two or more communities.}
\begin{eqnarray}
H_0\defi \sum_l H_0^{(l)}
-\sum_{l<k}\sum_{(i,j)\in\Gamma_0^{(l,k)}} J_{0;(i,j)}^{(l,k)}\sigma_{i}\sigma_{j},
\label{H0}
\end{eqnarray}
where the $h^{(l)}$ are arbitrary external fields and 
$J_{0;(i,j)}^{(l)}$'s and the $J_{0;(i,j)}^{(l,k)}$'s are arbitrary ``short-range'' couplings.
From now on, for shortness, we will use for them the simpler notations  
$J_{0}^{(l)}$'s and the $J_{0}^{(l,k)}$, respectively, as if they were uniform couplings. 
However, it should be kept in mind that there is no limitation in the choices of these
couplings, as well as in the choice of the graphs $(\mathcal{L}_0^{(l)},\Gamma_0^{(l)})$
and $(\mathcal{L}_0^{(l,k)},\Gamma_0^{(l,k)})$.

Let be given $n+n(n-1)/2$ independent random graphs $\bm{c}^{(l,k)}$, $l\leq k=1,\ldots,n$.
We will indicate by $c^{(l,l)}$ the average connectivity of the graph $\bm{c}^{(l,l)}$ 
(average with respect to a measure $P(\bm{c})$ we soon prescribe),
and by $c^{(l,k)}$ and $c^{(k,l)}$ the two directed average connectivities of the graph
$\bm{c}^{(l,k)}$ counting how many bonds, in the average, connect a given vertex of $\mathcal{L}_0^{(l)}$ 
with vertices of $\mathcal{L}_0^{(k)}$, and vice-versa, respectively.
Due to their definition, for $l\neq k$, $c^{(l,k)}$ and $c^{(k,l)}$ are not independent,
in fact it must hold the following balance equation
\begin{eqnarray}
\label{conn}
N^{(l)}c^{(l,k)}=N^{(k)}c^{(k,l)},
\end{eqnarray}
or else, by using (\ref{alpha})
\begin{eqnarray}
\label{conn1}
\alpha^{(l)}c^{(l,k)}=\alpha^{(k)}c^{(k,l)}.
\end{eqnarray}
Eq. (\ref{conn1}) suggests to define the following symmetric matrix
which we will soon use:
\begin{eqnarray}
\label{conn2}
\tilde{c}^{(l,k)}\defi \alpha^{(l)}c^{(l,k)},\quad \forall l,k=1,\ldots,n.
\end{eqnarray}
Besides the constrains (\ref{conn1}), it is important to recall that,
for finite $N$, the connectivities are bounded as follows
\begin{eqnarray}
\label{conn3}
0&\leq& c^{(l,k)} \leq \alpha^{(k)}N, 
\end{eqnarray}
or else, by using the symmetric matrix 
\begin{eqnarray}
\label{conn4}
0&\leq& \tilde{c}^{(l,k)} \leq \alpha^{(l)}\alpha^{(k)}N.
\end{eqnarray}

We will use the symbol $c_{i,j}^{(l,k)}$ to indicate the adjacency matrix elements of the graph $\bm{c}^{(l,k)}$: 
$c_{i,j}^{(l,k)}=0,1$, $i\in\mathcal{L}_0^{(l)},j\in\mathcal{L}_0^{(k)}$.
%$\bm{c}^{(l,k)}=\{c_{i,j}^{(l,k)}\}$.
The symbol $\bm{c}$ will indicate the graph obtained as union of all the
$n+n(n-1)/2$ graphs $\bm{c}^{(l,k)}$, $l\leq,k=1,\ldots,n$.

We now define our small-world models.
For each $l$ we super-impose the bonds of the random graph $\bm{c}^{(l,l)}$ 
to connect, through certain short-cuts, some vertices of $\mathcal{L}_0^{(l)}$, and for each couple $(l,k)$ we super-impose 
the bonds of the random graph $\bm{c}^{(l,k)}$ to connect, through certain short-cuts, some vertices of   
$\mathcal{L}_0^{(l)}$ with some vertices of $\mathcal{L}_0^{(l)}$, 
and define the corresponding small-world model on the $n$ communities 
- shortly \textit{the random model} - as described by the following Hamiltonian
\begin{eqnarray}
\label{H}
H_{\bm{c};\bm{J}}&\defi &H_0-\sum_l\sum_{~~i<j,~ i,j\in \mathcal{L}_0^{(l)}} 
c_{ij}^{(l,l)}{J}_{ij}^{(l,l)}\sigma_{i}\sigma_{j}
\nonumber \\
&& -\sum_{l<k}\sum_{~~i\in\mathcal{L}_0^{(l)},j\in\mathcal{L}_0^{(k)}} c_{ij}^{(l,k)}{J}_{ij}^{(l,k)}\sigma_{i}\sigma_{j},
\end{eqnarray}
the free energy $F$ and the averages $\overline{\media{\mathop{O}}^l}$, with $l=1,2$,
being defined in the usual (quenched) way as
%\begin{widetext}
\begin{eqnarray}
\label{logZ}
-\beta F\defi \sum_{\bm{c}} P(\bm{c})\int d\mathcal{P}
\left(\bm{J}\right)
\log\left(Z_{\bm{c};\bm{J}}\right),
\end{eqnarray} 
and 
\begin{eqnarray}
\label{O}
\overline{\media{\mathop{O}}^l}\defi 
\sum_{\bm{c}} P(\bm{c}) \int d\mathcal{P}\left(\bm{J}\right)
\media{\mathop{O}}^l, \quad l=1,2
\end{eqnarray} 
where $Z_{\bm{c};\bm{J}}$ 
is the partition function of the quenched system
\begin{eqnarray}
\label{Z}
Z_{\bm{c};\bm{J}}= \sum_{\{\sigma_{i}\}}
e^{-\beta H_{\bm{c};\bm{J}}\left(\{\sigma_i\}\}\right)}, 
\end{eqnarray} 
$\media{\mathop{O}}_{\bm{c};\bm{J}}$ the Boltzmann-average 
of the quenched system (note that $\media{\mathop{O}}_{\bm{c};\bm{J}}$ depends on the
given realization of the ${J}$'s and of $\bm{c}$:
$\media{\mathop{O}}=\media{\mathop{O}}_{\bm{c};\bm{J}}$;
for shortness we will often omit to write these dependencies)
\begin{eqnarray}
\label{OO}
\media{\mathop{O}}\defi \frac{\sum_{\{\sigma_i\}}\mathop{O}_{\bm{c};\bm{J}}e^{-\beta 
H_{\bm{c};\bm{J}}\left(\{\sigma_i\}\right)}}{Z_{\bm{c};\bm{J}}}, 
\end{eqnarray} 
and $d\mathcal{P}\left(\bm{J}\right)$ and $P(\bm{c})$ 
are two product measures given 
in terms of $n+n(n-1)/2$ normalized measures, $d\mu^{(l,k)}(J_{i,j}^{(l,k)})\geq 0$ 
and other $n+n(n-1)/2$ normalized measures $p^{(l,k)}(c_{i,j}^{(l,k)})\geq 0$, respectively:
%(we are considering a general measure $d\mu{i,j}$ 
%allowing also for a possible dependence on the bonds) 
\begin{eqnarray}
\label{dP}
d\mathcal{P}\left(\bm{J}\right)&\defi& \prod_l\prod_{~~i<j,~i,j\in\mathcal{L}_0^{(l)}} 
d\mu^{(l,l)}\left( {J}_{i,j}^{(l,l)} \right),
\nonumber \\
&\times&\prod_{l<k}\prod_{~~i\in\mathcal{L}_0^{(l)},j\in\mathcal{L}_0^{(k)}} 
d\mu^{(l,k)}\left( {J}_{i,j}^{(l,k)} \right),
\nonumber \\
&& \quad \int d\mu^{(l,k)}\left( {J}_{i,j}^{(l,k)} \right)=1,
\end{eqnarray}
\begin{eqnarray}
\label{Pg}
P(\bm{c})&\defi& \prod_l\prod_{~~i<j,~i,j\in\mathcal{L}_0^{(l)}} p^{(l,l)}(c_{i,j}^{(l,l)})
\nonumber \\
&\times&
\prod_{l<k}\prod_{~~i\in\mathcal{L}_0^{(l)},j\in\mathcal{L}_0^{(k)}} p^{(l,k)}(c_{i,j}^{(l,k)}),
\nonumber \\
&& \quad \sum_{c_{i,j}^{(l,k)}=0,1} p(c_{i,j}^{(l,k)})=1.
\end{eqnarray}
The variables
$c_{i,j}^{(l,k)}\in\{0,1\}$ specify whether a ``long-range'' bond between the sites
$i\in\mathcal{L}_0^{(l)}$ and $j\in\mathcal{L}_0^{(k)}$ 
is present ($c_{i,j}^{(l,k)}=1$) or absent ($c_{i,j}^{(l,k)}=0$), whereas
the $J_{i,j}^{(l,k)}$'s are the random couplings of the given bond $(i,j)$.
For $l\neq k$, the probability $p^{(l,k)}$ to select a bond connecting $\mathcal{L}_0^{(l)}$
with $\mathcal{L}_0^{(k)}$ among all the possible $N^{(l)}N^{(k)}$ bonds is given by 
$p^{(l,k)}=\tilde{c}^{(l,k)}/(N\alpha^{(l)}\alpha^{(k)})$. 
%However this equation holds also for $l=k$. 
Therefore the random variables $c_{i,j}^{(l,k)}$'s obey the following distributions
\begin{eqnarray}
\label{PP1}
 p(c_{ij}^{(l,k)})&=&
\frac{\tilde{c}^{(l,k)}}{N\alpha^{(l)}\alpha^{(k)}}\delta_{c_{ij}^{(l,k)},1}
\nonumber \\ 
&+& \left(1-\frac{\tilde{c}^{(l,k)}}{N\alpha^{(l)}\alpha^{(k)}}\right)\delta_{c_{ij}^{(l,k)},0},
\end{eqnarray}
which, for $l=k$ reduces to
\begin{eqnarray}
\label{PP}
 p(c_{ij}^{(l,l)})=
\frac{c^{(l,l)}}{N\alpha^{(l)}}
\delta_{c_{ij}^{(l,l)},1}+\left(1-\frac{c^{(l,l)}}{N\alpha^{(l)}}\right)\delta_{c_{ij}^{(l,l)},0}.
\end{eqnarray} 
Notice that the matrix entering Eq. (\ref{PP1}) is the symmetric one and not $c^{(l,k)}$,
%In the thermodynamic limit $N\to\infty$ each graph $\bm{c}^{(l,k)}$ will be distributed
%according to a Poisson law with the two directed average connectivities $c^{(l,k)}$ and $c^{(k,l)}$,
however, in the thermodynamic limit $N\to\infty$, for each $(l,k)$,
% for each graph $\bm{c}^{(l,k)}$
the degree random variables 
\begin{eqnarray}
\label{PP}
c_{i}^{(l,k)}\defi \sum_{j\in\mathcal{L}_0^{(k)}}c_{i,j}^{(l,k)}, \quad i\in\mathcal{L}_0^{(l)},
\end{eqnarray} 
will be distributed
according to a Poissonian law with the directed average connectivity $c^{(l,k)}$.

Concerning the measures $d\mu^{(l,k)}$, they are completely arbitrary.
When necessary, to be more specific in considering some example,
we shall assume one of the following typical measures
\begin{eqnarray}
\label{mea}
\frac{d\mu^{(l,k)}}{dJ_{i,j}^{(l,k)}}=\delta\left(J_{i,j}^{(l,k)}-J^{(l,k)}\right),
%\left(J_{i,j}^{(l,k)}\right)=\delta\left(J_{i,j}^{(l,k)}-J^{(l,k)}\right),
\end{eqnarray}
\begin{eqnarray}
\label{mea1}
\frac{d\mu^{(l,k)}}{dJ_{i,j}^{(l,k)}}=
\frac{1}{2}\delta\left(J_{i,j}^{(l,k)}-J^{(l,k)}\right)
+\frac{1}{2}\delta\left(J_{i,j}^{(l,k)}+J^{(l,k)}\right),
\end{eqnarray}
\begin{eqnarray}
\label{mea2}
\frac{d\mu^{(l,k)}}{dJ_{i,j}^{(l,k)}}=\sqrt{\frac{N}{2\pi\tilde{J}^{(l,k)}}}
\exp{\left[-\frac{\left(J_{i,j}^{(l,k)}-\frac{J^{(l,k)}}{N}\right)^2}{2\tilde{J}^{(l,k)}}N\right]},
\end{eqnarray}
where the parameters $J^{(l,k)}$ (not to be confused with the random variables $J_{i,j}^{(l,k)}$) are arbitrary, and $\tilde{J}^{(l,k)}>0$.

%A fundamental concept in antiferromagnetism is partitioning.
%We recall that a lattice $\mathcal{L}_0$ with bonds $\Gamma_0$ is said to be bipartite if 
%we can divide $\mathcal{L}_0$ into two parts such that all the given bonds only connect 
%one part with the other. For later use, it is convenient to translate this definition
%formally as follows.
%The couple $(\mathcal{L}_0,\Gamma_0)$
%is said to be bipartite if there exist two sublattices 
%$\mathcal{L}_0^{(A)},\mathcal{L}_0^{(B)}\subset \mathcal{L}_0$ and two subsets of bonds
%$\Gamma_0^{(A/B)},\Gamma_0^{(B/A)}\subset \Gamma_0$ such that
%\begin{eqnarray}
%\label{PP}
%&& \mathcal{L}_0^{(A)}\cup\mathcal{L}_0^{(B)}=\mathcal{L}_0, \quad \mathcal{L}_0^{(A)}\cap\mathcal{L}_0^{(B)}=\emptyset, 
%\nonumber \\
%&& \Gamma_0^{(A/B)}\cup\Gamma_0^{(B/A)}=\Gamma_0, \quad \Gamma_0^{(A/B)}\cap\Gamma_0^{(B/A)}=\emptyset, 
%\nonumber \\
%&& \mathrm{s.t.:}~ \mathrm{if}~ (i,j)\in \Gamma_0^{(X/Y)},~ \mathrm{then}
%\nonumber \\
%&& i\in \mathcal{L}_0^{(X)}, j\in\mathcal{L}_0^{(Y)} 
%~\mathrm{or}~ j\in \mathcal{L}_0^{(X)}, i\in\mathcal{L}_0^{(Y)},  
%%\end{eqnarray}
%where $X$ and $Y$ can be either $A$ or $B$.
%More in general, the couple $(\mathcal{L}_0,\Gamma_0)$ is said to be an $n$-partite graph
%if $\mathcal{L}_0$ can be splitted in $n$ subalattices so that any vertex on a given 
%sublattice interact with, and only with, vertices belonging to the other $n-1$ sublattices
%\cite{Bipart}.

\section{An effective field theory}
\subsection{The self-consistent equations for $n$ communities}
Physically, depending on the temperature $T$, and 
on the parameters of the probability distributions $\{d\mu^{(l,k)}\}$ and the connectivities $\{c^{l,k}\}$,
the random model may stably stay either in the phases P, F, SG, or AF. 
However, as we have already showed in the Ref. \cite{SW}, in our approach 
for any choice of $T$, $d\mu$ and $c$, independently of the signs of the couplings
and on the fact that the corresponding order parameters are zero or not, 
for the free energy, and then for any observable, there are always two - and only two - stable solutions
that we label as F and SG and that in the thermodynamic limit only one
of the two survives. Therefore, an AF like phase in our approach is not represented
by another solution; an AF like phase, if any, occurs in the solution with label F.   
In the Ref. \cite{SW} we showed that, for $n=1$, for the solution with label F and SG 
there are two natural decoupled order parameters that we have indicated 
as $m^{(\mathrm{F})}$ and $m^{(\mathrm{SG})}$, respectively.
Similarly, now we have $n$ coupled order parameters $m^{(\mathrm{F};l)}$ for the solution F 
and $n$ other coupled order parameters $m^{(\mathrm{SG};l)}$ for the solution SG, $l=1,\ldots,n$.
All the results we provide are exact up to $\mathop{O}(1/N)$ corrections.
%Similarly, if the couple $(\mathcal{L}_0,\Gamma_0)$ is a bipartite graph, 
%for an AF phase there are two natural order parameters $m^{(\mathrm{A})}$ and $m^{(\mathrm{B})}$ 
%giving the magnetizatios on the sublattices $\mathcal{L}_0^{(A)}$ and $\mathcal{L}_0^{(B)}$, respectively.
%In the following, we will use the label $\mathop{}_0$ 
%to specify that we are referring
%to the pure model with Hamiltonian (\ref{H0}).
%Note that all the equations presented in this paper 
%have meaning and usefulness also for sufficiently large but
%finite size $N$. For shortness we shall omit to write the dependence on $N$.
%Let $m_0^{(A)}(\beta J_0;\beta h^{(A)},\beta h^{(B)})$ and $m_0^{(B)}(\beta J_0;\beta h^{(A)},\beta h^{(B)})$ 
%be the stable magnetizations on $\mathcal{L}_0^{(A)}$ and $\mathcal{L}_0^{(B)}$, respectively,
%of the pure model with coupling $J_0$ and in the presence of two  
%uniform external fields $h^{(A)}$ and $h^{(B)}$ acting on $\mathcal{L}_0^{(A)}$ and $\mathcal{L}_0^{(B)}$.
%In other words, $m_0^{(A)}$ and $m_0^{(B)}$ are the Gibbs averages for spins on  
%$\mathcal{L}_0^{(A)}$ and $\mathcal{L}_0^{(B)}$ for the following pure Hamiltonian
%\begin{widetext}
%\begin{eqnarray}
%H_0^{(A,B)}&\defi& -J_{0}\sum_{(i,j)\in \Gamma_0}\sigma_{i}\sigma_{j} 
%%\nonumber \\ && 
%-h^{(A)}\sum_{i\in\mathcal{L}_0^{(A)}} \sigma_i-h^{(B)}\sum_{i\in\mathcal{L}_0^{(B)}} \sigma_i.
%\label{H0AB}
%\end{eqnarray} 
\subsubsection{$J_0^{(l,k)}=0$ for $l\neq k$}
Let us consider the interesting case in which there are no short-range 
interactions between different communities, \textit{i.e.}, let us assume that $J_0^{(l,k)}=0$ for $l\neq k$.
Let $m_0^{(l)}(\beta J_0^{(l)};\beta h^{(l)})$ be the stable magnetization of the pure model
with Hamiltonian (\ref{H0l}). 
Then, for both $\Sigma=$F and SG, the $n$ order parameters $m^{(\Sigma;l)}$
satisfy independently the following system of $n$ coupled equations
\begin{eqnarray}
\label{THEO}
m^{(\Sigma;l)}=m_0^{(l)}\left(\beta J_0^{(\Sigma;l)}; \beta H^{(\Sigma;l)}+\beta h^{(l)}\right),
%\quad l=1,\ldots,n,
\end{eqnarray}
where
\begin{eqnarray}
\label{THEO1}
\beta H^{(\Sigma;l)}\defi \sum_k \beta J^{(\Sigma;l,k)}m^{(\Sigma;k)},
%\quad l=1,\ldots,n,
\end{eqnarray}
and
%, if $i\in\mathcal{L}_0^{(l)}$ and  $j\in\mathcal{L}_0^{(k)}$,
the effective couplings $J^{(\mathrm{F};l,k)}$, $J^{(\mathrm{SG};l,k)}$,
$J_0^{(\mathrm{F};l)}$ and $J_0^{(\mathrm{SG};l)}$ are given by
%~\footnote{Notice that
%with respect to the definitions (14)-(17) of part I we have here put out the connectivities.}
\begin{eqnarray}
\label{THEO2}
\beta J^{(\mathrm{F};l,k)}\defi c^{(l,k)}\int d\mu^{(l,k)}(J_{i,j}^{(l,k)})\tanh(\beta J_{i,j}^{(l,k)}),
\end{eqnarray} 
\begin{eqnarray}
\label{THEO3}
\beta J^{(\mathrm{SG};l,k)}\defi c^{(l,k)}\int d\mu^{(l,k)}(J_{i,j}^{(l,k)})\tanh^2(\beta J_{i,j}^{(l,k)}),
\end{eqnarray} 
\begin{eqnarray}
\label{THEO4}
J_0^{(\mathrm{F};l)}\defi J_0^{(l)},
\end{eqnarray} 
and
\begin{eqnarray}
\label{THEO5}
\beta J_0^{(\mathrm{SG};l)}\defi \tanh^{-1}(\tanh^2(\beta J_0^{(l)})).
\end{eqnarray}
Note that, when $\alpha^{(l)}\neq \alpha^{(k)}$, unlike the random couplings $J_{i,j}^{(l,k)}$, the
effective couplings $J^{(\mathrm{F};l,k)}$ and $J^{(\mathrm{SG};l,k)}$ are not symmetric.
However, as we shall see soon, the couplings entering the free energy are
the symmetric ones: $\alpha^{(l)}J^{(\Sigma;l,k)}$. 
Note also that $|J_0^{(\mathrm{F};l)}|>J_0^{(\mathrm{SG};l)}$.

For a correlation function $C_r^{(\Sigma;l)}$ of degree $r$, 
involving a set of $r$ vertices belonging only to the same community $\mathcal{L}_0^{(l)}$ we have 
%up to $\mathop{O}\left(1/N\right)$ corrections we have 
\begin{eqnarray}
\label{THEO6}
{{C}}_r^{(\Sigma;l)}&=&
{{C}}_{0r}^{(l)}\left(\beta J_0^{(\Sigma;l)};\beta H^{(\Sigma;l)}+\beta h^{(l)}\right),
%\nonumber \\
%+\mathop{O}\left(\frac{1}{N}\right),
\end{eqnarray} 
whereas for a correlation function $C_{r,s}^{(\Sigma;l,k)}$ of degree $r+s$, involving 
a set of $r$ vertices belonging to $\mathcal{L}_0^{(l)}$ and 
a set of $s$ vertices belonging to $\mathcal{L}_0^{(k)}$, with $l\neq k$, we have
\begin{eqnarray}
\label{THEO7}
{{C}}_{r,s}^{(\Sigma;l,k)}&=&%{{C}}_{0r}^{(\Sigma;l)}{{C}}_{0s}^{(\Sigma;k)}
{{C}}_0^{(l)}\left(\beta J_0^{(\Sigma;l)};\beta H^{(\Sigma;l)}+\beta h^{(l)}\right)
\nonumber \\
&\times& {{C}}_0^{(k)}\left(\beta J_0^{(\Sigma;k)};\beta H^{(\Sigma;k)}+\beta h^{(k)}\right),
%\nonumber \\
%+\mathop{O}\left(\frac{1}{N}\right),
\end{eqnarray}  
where the ${{C}}_{0r}^{(l)}(\beta J_0^{(l)};\beta h^{(l)})$'s  
are the corresponding correlation functions of degree $r$ of the pure 
model with Hamiltonian (\ref{H0l}).
For the specific relation between the above correlation functions and the averages or quadratic averages 
of physical observables, we remind the reader to Eqs. (24)-(28) of the Ref. \cite{SW}.
In particular in the F phase we have
\begin{eqnarray}
\label{THEO8}
\overline{\media{\sigma_i}}=m^{(\mathrm{F};l)}, \quad i\in\mathcal{L}_0^{(l)},
\end{eqnarray}  
from which, by using (\ref{alpha}) and (\ref{alpha1}), 
it follows also that the average magnetization $m^{(\mathrm{F})}$ 
is given by~\footnote{If instead of the normalization (\ref{alpha1}) we leave
the $\alpha$'s arbitrary, the correlation functions must be divided by the sum of the $\alpha$'s;
as $m^{(\mathrm{F})}=\frac{\sum_l \alpha^{(l)}m^{(\mathrm{F};l)}}{\sum_l \alpha^{(l)}}$.}
\begin{eqnarray}
\label{THEO9}
m^{(\mathrm{F})}=\sum_l \alpha^{(l)}m^{(\mathrm{F};l)}, 
\end{eqnarray}  
similarly in the SG phase we have 
\begin{eqnarray}
\label{THEO8b}
\overline{\media{\sigma_i}^2}={m^{(\mathrm{SG};l)}}^2, \quad i\in\mathcal{L}_0^{(l)},
\end{eqnarray}  
\begin{eqnarray}
\label{THEO9b}
{m^{(\mathrm{SG})}}^2=\sum_l \alpha^{(l)}{m^{(\mathrm{SG};l)}}^2.
\end{eqnarray}   

The free energy density $f^{(\Sigma)}$ coming from Eq. (\ref{logZ}) involves a generalized Landau free energy density $L^{(\Sigma)}$ from
which it differs only for trivial terms independent from the $m^{(\Sigma)}$'s.
The complete expression for $f^{(\Sigma)}$ in terms of $L^{(\Sigma)}$ is left to the reader and corresponds to 
the obvious generalization of Eq. (21) of the Ref. \cite{SW}. 
The term $L^{(\Sigma)}$ reads ($\beta f^{(\Sigma)}=$ trivial term $+L^{(\Sigma)}/l^{(\Sigma)}$, with $l^{(\Sigma)}=1,2$
for $\Sigma=$F, SG, respectively)
\begin{widetext}
\begin{eqnarray}
\label{THEO10}
&& L^{(\Sigma)}\left(m^{(\Sigma;1)},\ldots,m^{(\Sigma;n)}\right)=\sum_l\alpha^{(l)}\beta g^{(\Sigma;l)},
\end{eqnarray} 
where 
\begin{eqnarray}
\label{THEO11}
\beta g^{(\Sigma;l)}\defi 
\frac{m^{(\Sigma;l)}}{2}\beta H^{(\Sigma;l)}+\beta f_0^{(l)}(\beta J_0^{(l)};\beta H^{(\Sigma;l)}+\beta h^{(l)}),
\end{eqnarray} 
$f_0^{(l)}(\beta J_0^{(l)},\beta h^{(l)})$ being the free energy density in the thermodynamic
limit of the pure model with Hamiltonian (\ref{H0l}). 
Eq. (\ref{THEO10}) can be also expressed in a more symmetric way as
\begin{eqnarray}
\label{THEO12}
L^{(\Sigma)}\left(m^{(\Sigma;1)},\ldots,m^{(\Sigma;n)}\right)&=&
\sum_{l, k} \frac{\alpha^{(l)}\beta J^{(\Sigma;l,k)}m^{(\Sigma;l)}m^{(\Sigma;k)}}{2}
+\sum_l\alpha^{(l)}\beta f_0^{(l)}\left(\beta J_0^{(l)};\beta H^{(\Sigma;l)}+\beta h^{(l)}\right).
\end{eqnarray} 
%however Eq. (\ref{THEO11}) reveals the peculiarity of the case
%$J_0^{(l,k)}=0$ for $l\neq k$. In fact, Eqs. (\ref{THEO10})-(\ref{THEO11}) say that the
%set of the effective fields $H^{(\Sigma;l)}$ 
%is a set of normal modes through which we can decouple analytically the self consistent Eq. (\ref{THEO}).
%Due to the non linearity of Eq. (\ref{THEO}), peforming such a decopuling is a non trivial task,
%however is always possible and, as we will see in Sec. IIIB4, 
%the existence of normal modes have important consequences.
%%\begin{eqnarray}
%%\label{THEO12}
%%L^{(\Sigma)}\left(m^{(\Sigma;1)},\ldots,m^{(\Sigma;n)}\right)&=&\sum_l\alpha^{(l)}
%%\left[\frac{\beta J^{(\Sigma;l,l)}\left(m^{(\Sigma;l)}\right)^2}{2}+
%%  \beta f_0^{(l)}(\beta J_0^{(l)};\beta H^{(\Sigma;l)}+\beta h^{(l)})\right]
%%\nonumber \\
%%&+& \sum_{l\neq k} \frac{\alpha^{(l)}\beta J^{(\Sigma;l,k)}m^{(\Sigma;l)}m^{(\Sigma;k)}}{2}.
%%\end{eqnarray} 

\subsubsection{$\exists (l,k)$, with $l\neq k$, such that $J_0^{(l,k)}\neq 0$}
Here we consider the most general case in which
there are at least two communities that 
interact also via short-range couplings. 
Now the additivity of the free energy of the pure model with respect to the communities
is completely lost and for any $l$ we need to consider in general
$m_0^{(l)}(\{\beta J_0^{(l',k')}\};\{\beta h^{(l')}\})$, the stable magnetization of the $l$-th community of the pure model
with the total Hamiltonian (\ref{H0}) having, in general, $n+n(n-1)/2$ short-range couplings 
$\{\beta J_0^{(l',k')}\}$ and $n$ external fields $\{\beta h^{(l')}\}$
(we use the parenthesis $\{\cdot\}$ as a short notation to indicate a vector or a matrix with 
components $l'=1,\ldots,n$ or $l',k'=1,\ldots,n$, respectively), 
where we have also introduced $J_0^{(l,l)}\defi J_0^{(l)}$. 
Then, the order parameters $m^{(\Sigma;l)}$, for both $\Sigma=$F and SG,   
satisfy the following system of $n$ coupled equations
\begin{eqnarray}
\label{THEOG}
m^{(\Sigma;l)}=m_0^{(l)}\left(\left\{\beta J_0^{(\Sigma;l',k')}\right\}; 
\left\{\beta H^{(\Sigma;l')}+\beta h^{(l')}\right\}\right),
%\quad l=1,\ldots,n,
\end{eqnarray}
where the effective fields $H^{(\Sigma;l)}$ and the effective couplings are defined as in 
Eqs. (\ref{THEO1})-(\ref{THEO5}) (with the obvious generalization for $J_0^{(\Sigma;l,k)}$).

For a correlation function $C_r^{(\Sigma;l)}$ of degree $r$, 
involving a set of $r$ vertices belonging to the same $l$-th community we have again the obvious
generalization of (\ref{THEO6}), while the obvious generalization of
Eq. (\ref{THEO7}) can hold
only between two groups of communities, say with index $l$ and $k$,
having no short-range couplings: $J_0^{(l,k)}=0$.
Eqs. (\ref{THEO8}) and (\ref{THEO9}) of course still hold,
while the term $L^{(\Sigma)}$ now becomes
\begin{eqnarray}
\label{THEO12G}
L^{(\Sigma)}\left(m^{(\Sigma;1)},\ldots,m^{(\Sigma;n)}\right)&=&
\sum_{l,k} \frac{\alpha^{(l)}\beta J^{(\Sigma;l,k)}m^{(\Sigma;l)}m^{(\Sigma;k)}}{2}
+ \beta f_0\left(\left\{\beta J_0^{(l',k')}\right\};\left\{\beta H^{(\Sigma;l')}+\beta h^{(l')}\right\}\right),
\end{eqnarray} 
%\begin{eqnarray}
%\label{THEO12G}
%L^{(\Sigma)}\left(m^{(\Sigma;1)},\ldots,m^{(\Sigma;n)}\right)&=&\sum_l\alpha^{(l)}
%\frac{\beta J^{(\Sigma;l,l)}\left(m^{(\Sigma;l)}\right)^2}{2}+
%\sum_{l\neq k} \frac{\alpha^{(l)}\beta J^{(\Sigma;l,k)}m^{(\Sigma;l)}m^{(\Sigma;k)}}{2}
%\nonumber \\
%&+& \beta f_0\left(\left\{\beta J_0^{(l',k')}\right\};\left\{\beta H^{(\Sigma;l')}+\beta h^{(l')}\right\}\right),
%\end{eqnarray} 
$f_0\left(\left\{\beta J_0^{(l',k')}\right\},\left\{\beta h^{(l')}\right\}\right)$ 
being the free energy density in the thermodynamic
limit of the pure model with the total Hamiltonian (\ref{H0}). 
When $J_0^{(l',k')}=0$ for any $l\neq k$, Eqs. (\ref{THEOG}) and (\ref{THEO12G}) reduce  
to Eqs. (\ref{THEO}) and (\ref{THEO12}), respectively. 

For given $\beta$, among all the possible solutions of the self-consistent system (\ref{THEO})-(\ref{THEO1})
(or (\ref{THEOG})), 
whose set we indicate by $\mathcal{M}$,
in the thermodynamic limit, for both $\Sigma$=F and $\Sigma$=SG, 
the true solution $\left(\bar{m}^{(\Sigma;1)},\ldots,\bar{m}^{(\Sigma;n)}\right)$, or leading solution, 
is the one that minimizes $L^{(\Sigma)}$:
\begin{eqnarray}
\label{THEO13}
&& L^{(\Sigma)}\left(\bar{m}^{(\Sigma;1)},\ldots,\bar{m}^{(\Sigma;n)}\right)=%\nonumber \\ &&
%\min_{(m^{(1)},\ldots,m^{(n)})\in [-1,1]^n} L^{(\Sigma)}\left(m^{(1)},\ldots,m^{(n)}\right).
\min_{\left(m^{(\Sigma;1)},\ldots,m^{(\Sigma;n)}\right)\in \mathcal{M}} L^{(\Sigma)}\left(m^{(\Sigma;1)},\ldots,m^{(\Sigma;n)}\right).
\end{eqnarray} 
\end{widetext}

For the localization and the reciprocal stability between the F and the SG phases  
see the discussion in Sec. IIID.

As an immediate consequence of the Eq. (\ref{THEOG}) we get that the adimensional susceptibility of the random model, 
$\tilde{\chi}^{(l,k)}\defi\partial m^{(\Sigma;l)}/\partial (\beta h^{(k)})$, written in matrix form is 
\begin{eqnarray}
\label{THEOsusc}
\bm{\tilde{\chi}^{(\Sigma)}}=\left(\bm{1}-\bm{\tilde{\chi}_0^{(\Sigma)}}\cdot\bm{\beta J^{(\Sigma)}}\right)^{-1}
\cdot \bm{\tilde{\chi}_0},
\end{eqnarray}
where we have introduced the matrix of the effective long-range couplings
$\beta J^{(\Sigma;l,k)}$, and $\tilde{\chi}_0^{(l,k)}\left(\beta J_0^{(l)};\beta h^{(l)}\right)$,
the adimensional susceptibility of the $l$-th pure community with respect
to the external field $h^{(k)}$ of the $k$-th community:
\begin{eqnarray}
\label{THEOh}
\tilde{\chi}_0^{(l,k)}\defi 
\frac{\partial m_0^{(l)}\left(\left\{\beta J_0^{(\Sigma;l',k')}\right\};\left\{\beta h^{(l')}\right\}\right)}
{\partial \beta h^{(k)}}.
\end{eqnarray}
Note that in the case $J_0^{(l,k)}=0$ for $l\neq k$, $\bm{\tilde{\chi}_0^{(\Sigma)}}$ is a diagonal matrix 
whereas $\bm{\tilde{\chi}^{(\Sigma)}}$ not.\\
 
\textit{Remark 1.}\label{Remark1} 
Note that, as it will become clear by looking at the proof in Sec. VII,
unlike the case $n=1$, for $n\geq 2$ the expression of $L^{(\Sigma)}$ 
in Eqs. (\ref{THEO10}) or (\ref{THEO12G}) has a physical meaning only when calculated at 
any stable solution of the self-consistent system (\ref{THEO}) or (\ref{THEOG}), respectively.
In fact, for $n\geq 2$, the free energy term $L^{(\Sigma)}$ is different
from the original density functional of the model $\mathcal{L}^{(\Sigma)}$ that lives in an enlarged space
of the order parameters; the form of $\mathcal{L}^{(\Sigma)}$ is equal to the form of $L^{(\Sigma)}$ 
only when calculated in a solution of the self-consistent system of Eqs. (\ref{THEO10}) or (\ref{THEO12G}).
In this sense, for $n\geq 2$, the expression ``Landau free energy density''
for $L^{(\Sigma)}$ would be somehow inappropriate; the true Landau free energy density is
represented by $\mathcal{L}^{(\Sigma)}$ and is given in Sec. VII, but unfortunately
its general expression turns out to be too complicated to be exploited for calculating
rigorously the stability of a given solution. We shall come back soon on this point in the next Section.
We stress however that Eqs. (\ref{THEO10}) or (\ref{THEO12G}) cover all the solutions,
stable or not; in other words at any saddle point (see Sec. VII) of the original 
density functional $\mathcal{L}^{(\Sigma)}$ we have $L^{(\Sigma)}=\mathcal{L}^{(\Sigma)}$.\\ 

\subsection{Stability and phase transition's scenario}
Note that, for $\beta$ sufficiently small (see later) and $\left\{h^{(l)}\right\}=0$, 
Eqs. (\ref{THEO}) or (\ref{THEOG}) have always the solution $\left\{m^{(\Sigma;l)}\right\}=0$ and, furthermore,
if $\left\{m^{(\Sigma;l)}_+\right\}$ is a solution for $\left\{h^{(l)}\right\}=0$, 
$\left\{m^{(\Sigma;l)}_-\right\}\defi -\left\{m^{(\Sigma;l)}_+\right\}$ is a solution as well. 
From now on, if not explicitly said, 
we will refer only to one of the dual solutions.
Equations (\ref{THEO}) or (\ref{THEOG}) define
a $n$ dimensional map.
Under this map a solution $\left\{m^{(\Sigma;l)}\right\}$ of Eq. (\ref{THEO}) or (\ref{THEOG}) 
is stable (but in general not unique) if
\begin{eqnarray}
\label{THEOs}
|\lambda_l|<1, \quad l=1,\ldots,n
\end{eqnarray}
where $\{\lambda_l\}$ are the eigenvalues of the $n\times n$ matrix
$\bm{\tilde{\chi}_0^{(\Sigma)}}\cdot\bm{\beta J}^{(\Sigma)}$ calculated at zero field:
$m^{(\Sigma;l)}=0,~~l=1,\ldots,n$.\\

\textit{Remark 2.}\label{Remark2}
Given a solution of the self-consistent system (\ref{THEO}) or (\ref{THEOG}),
Eq. (\ref{THEOs}) represents the stability condition of the solution under iteration
as an $n$-dimensional map. As we have mentioned in \textit{Remark 1}, due to the fact that, for $n\geq 2$, the original 
density functional of the model $\mathcal{L}^{(\Sigma)}$ and the term $L^{(\Sigma)}$
are different, we cannot calculate the true Hessian of $\mathcal{L}^{(\Sigma)}$ from
the Hessian of $L^{(\Sigma)}$. Unfortunately, the Hessian of $\mathcal{L}^{(\Sigma)}$
has a quite complicated form even when calculated at a solution 
of the self-consistent system (\ref{THEO}) or (\ref{THEOG}). The positivity of this Hessian  
would be important to discriminate rigorously the stability of any solution. 
In this sense, as done for $n=1$ in the Ref. \cite{SW},
when the transition is of second-order, 
important information about the critical behavior of the system 
could be obtained by expanding $\mathcal{L}^{(\Sigma)}$ around the solution $\left\{m^{(\Sigma;l)}\right\}=0$ 
by keeping a sufficiently
large number of terms involving the even derivatives of the matrix 
$\bm{A}^{(\Sigma)}\left(\left\{\beta J_0^{(l,k)}\right\};\left\{\beta H^{(\Sigma;l)}+\beta h^{(l)}\right\}\right)$
with respect to the external fields $\beta h^{(l)}$ and calculated at
$\left\{\beta H^{(\Sigma;l)}\right\}=\left\{\beta h^{(l)}\right\}=0$. 
Such a general study is beyond the aim of this paper. 
Note however that, even if we are not able to discriminate rigorously between stable and unstable states,
due to the fact that the self-consistent system (\ref{THEO}) or (\ref{THEOG})
give all the possible solutions, for any given $\beta$ we are
able to predict exactly which is the leading (and then also stable) solution by
looking at the solution that gives the absolute minimum of $L^{(\Sigma)}$,
even when there are first-order phase transitions~\footnote{Of course we make the natural physical
assumption that when the saddle point equation admits only one solution this be a stable
solution.}.\\ 

By setting $\left\{h^{(l)}\right\}=0$ and expanding Eq. (\ref{THEO}) or (\ref{THEOG}) to the first order,
we get the equation for the critical temperature
$1/\beta_c^{(\Sigma)}$ of a P-$\Sigma$ phase transition when it is of second-order:
\begin{eqnarray}
\label{THEOs2al}
\max\limits_{l=1,\ldots,n}|\lambda_l|=1,
\end{eqnarray}
which implies
\begin{eqnarray}
\label{THEOs2}
\det \left(\bm{A}^{(\Sigma)}\left(\left\{\beta_c^{(\Sigma)} J_0^{(\Sigma;l,k)}\right\};\left\{0\right\}\right)\right)=0,
\end{eqnarray}
where the $n\times n$ matrix $\bm{A}^{(\Sigma)}$ is given by
\begin{eqnarray}
\label{THEOs1}
\bm{A}^{(\Sigma)}\defi \bm{1}-\bm{\tilde{\chi}_0^{(\Sigma)}}\cdot\bm{\beta J}^{(\Sigma)}|_{m^{(\Sigma;l)}=0,~~l=1,\ldots,n}.
\end{eqnarray}
%\begin{eqnarray}
%\label{THEOs1}
%&& A^{(\Sigma ;l,k)}=%\left(\left\{\beta J_0^{(l)}\right\};\left\{\beta H^{(l)}+\beta h^{(l)}\right\}\right)=
%\delta_{l,k} - \beta J^{(\Sigma;l,k)}\times
%\nonumber\\
%&\times&\tilde{\chi}_0^{(l,l)}\left(\beta J_0^{(\Sigma;l)};\beta H^{(\Sigma;l)}+\beta h^{(l)}\right),
%\end{eqnarray}
%if $\beta J_0^{(l,k)}=0$ for any $l\neq k$, whereas in the general case
%\begin{eqnarray}
%\label{THEOs1G}
%&& A^{(\Sigma ;l,k)}=%\left(\left\{\beta J_0^{(l)}\right\};\left\{\beta H^{(l)}+\beta h^{(l)}\right\}\right)=
%\delta_{l,k} - \sum_m\beta J^{(\Sigma;m,k)} \times \nonumber \\
%&\times& \tilde{\chi}_0^{(l,m)}\left(\left\{\beta J_0^{(\Sigma;l',k')}\right\};\left\{\beta H^{(\Sigma;l')}+\beta h^{(l')}\right\}\right).
%\end{eqnarray}

Eqs. (\ref{THEOs2al}) or (\ref{THEOs2}) generalize Eq. (44) of the Ref. \cite{SW} to which reduce for $n=1$.
In the Ref. \cite{SW} we have seen that: when $J_0\geq 0$ (and then $J_0^{(\Sigma)}\geq 0$), 
independently of the added connectivity $c$ and independently of the sign of the shortcuts $J$'s
(and then independently of the sign of $J^{(\Sigma)}$), the phase transition, both P-F or P-SG, 
is second-order and the critical indices are
the classic ones; while, when $J_0<0$, we still have $J_0^{(\mathrm{SG})}\geq 0$
an then the P-SG transition is still second-order but, due to the fact that
now $J_0^{(\mathrm{F})}<0$, for a sufficiently large $c$,
there are multicritical points and, moreover,
it may appear P-F first-order phase transitions and in such a case 
the critical temperature in general does not satisfy Eq. (\ref{THEOs2}) and the critical behavior
can belong to the so called $m^l$ theory of Landau phase transitions, $l$ being an even 
integer greater or equal to 6.
However, when $n\geq 2$, %unless the shortcuts $J$'s have positive averages, 
the above ``simpler'' dual scenario 
for $J_0\geq 0$ and $J_0<0$ is in general no more valid. 
In fact, if for $n=1$ in Eq. (\ref{THEOs}) we set $\beta J^{(\mathrm{F})}<0$,
we see that the solution $m^{(\mathrm{F})}=0$ is always stable (recall that the susceptibility is always non negative), 
while, if - for $n\geq 2$ - for some couple $(l,k)$ with $l\neq k$, 
in Eq. (\ref{THEOs}) we set $\beta J^{(\mathrm{F};l,k)}<0$,
we see that in general the solution $\left\{m^{(\mathrm{F};l)}\right\}=0$ is no more
a stable solution, even if $J_0^{(l,l)}\geq 0$ and $J_0^{(l,k)}=0$ for any $l\neq k$
(try for example the simplest case: $n=2$, $J_0^{(l,k)}=0$ and $J^{(\Sigma;1,1)}=J^{(\Sigma;2,2)}=0$). 
This effect is of course at the base
of antiferromagnetism %(we shall provide important examples later)
and gives a clue on how much more complex will be now
the scenario of phase transitions, be P-F or P-SG like.

According to the symmetries of the effective couplings,  
we distinguish three cases: the homogeneous case, the symmetric case
and the general case.

\subsubsection{The homogeneous case and the absence of internal antiferromagnetism} 
Let us consider the uniform case, \textit{i.e.}, the case in which 
the $n$ communities have: equal size, $\alpha^{(l)}=1/n$, 
equal connectivity, $c^{(l,k)}=c$, and interact through: arbitrary short-range couplings,
$\{J_0^{(l,k)}\}$, but through equally distributed long-range couplings,
$d\mu^{(l,k)}=d\mu$ also for $l=k$. Therefore, these models have only one
effective long-range coupling, say $\beta J^{(\Sigma)}$, and only one order parameter, say $m^{(\Sigma)}$, and
reduce to the small-world models of one community already studied in the Ref. \cite{SW} whose self-consistent equation,
in its most general form to include $n$ arbitrary external fields, was given by (from Eqs. (A7)-(A12) of the Ref. \cite{SW})
\begin{widetext}
\begin{eqnarray}
\label{App7}
{{m}}^{(\Sigma)}=\frac{1}{n}\sum_lm_{0}^{(l)}
\left(\left\{\beta J_0^{(\Sigma;l',k')}\right\},\left\{\beta J^{(\Sigma)}{{m}}^{(\Sigma)}n + \beta h^{(k')}\right\}\right),
\end{eqnarray}
\begin{eqnarray}
\label{App10}
m^{(\Sigma;l)}=m_{0}^{(l)}
\left(\left\{\beta J_0^{(\Sigma;l',k')}\right\},\left\{\beta J^{(\Sigma)}{{m}}^{(\Sigma)}n + \beta h^{(k')}\right\}\right),
\end{eqnarray}
\begin{eqnarray}
\label{App11}
m^{(\Sigma)}=\frac{1}{n}\sum_l m^{(\Sigma;l)},
\end{eqnarray}
\end{widetext}
where we have used the definitions (\ref{THEO2})-(\ref{THEO5}) 
and we have taken into account that the total average connectivity $c_t$ seen by each community is $c_t=c n$.
Our general solution, Eq. (\ref{THEOG}), reproduces - of course - this
result, but it is interesting to observe that this effect can be seen as due 
a particular case of the super-position principle that
emerges in our self-consistent equations for the general problem.
Note that in this special case, despite the existence of $n$
communities, for each phase F or SG, there is just one order parameter $m^{(\Sigma)}$.
This fact implies serious limitations on the possible phases of such a model. 
In fact, let us consider for simplicity the case in which all the communities
have the same internal short-range coupling $J_0^{(l,l)}=J_0$ 
and suppose also that there is no short-range coupling among
different communities: $J_0^{(l,k)}=0$ for $l\neq k$. 
From Eqs. (\ref{App7})-(\ref{App11})% and from the stability condition (\ref{THEOs}), 
we see that for $\{h^{(l)}\}=\{0\}$, and independently of $J_0$,
if $J^{(\mathrm{F})}>0$, all the $m^{(\mathrm{F};l)}$ are parallel 
(recall that, for $h\neq 0$, at equilibrium the sign of the thermal average magnetization 
of the single $l$-th community, $m_0^{(l)}(\beta J_0;\beta h)$, is equal to the sign of $h$).
If instead $J^{(\mathrm{F})}<0$, at any temperature the only stable solution of Eq. (\ref{App7}) is
$m^{(\mathrm{F})}=0$, 
and since the communities do not interact (from the point of view of our effective field theory), 
from Eq. (\ref{App7}) it also follows that
$\{m^{(\mathrm{F};l)}\}= \{0\}$. 
More in general, this result holds essentially also when
we allow for the presence of a same short-range coupling among different communities,
say: $J_0^{(l,k)}=J_0^{(1,2)}$, for $l\neq k$. In fact in this case we 
have that for $\{h^{(l)}\}=\{0\}$, if $J_0^{(1,2)}\geq 0$ and
if $J^{(\mathrm{F})}>0$, all the $m^{(\mathrm{F};l)}$ tend to be parallel and equal
to the - single - order parameter $m^{(\mathrm{F})}$; whereas,
if $J^{(\mathrm{F})}<0$, at any temperature the only stable solution of Eq. (\ref{App7}) is
again $m^{(\mathrm{F})}=0$ which, in turn, implies that, 
due to Eq. (\ref{App11}), we must have also $\{m^{(\mathrm{F};l)}\}=\{0\}$. 
A similar argument for $J^{(\mathrm{F})}<0$ cannot however be repeated if some
of the short-range couplings $J_0^{(l,k)}$ are negative. In such a case in fact, 
below the critical temperature $T_{AF;0}$ - if any - 
of a possible antiferromagnetic phase transition of the pure model, 
the pure magnetizations $m_0^{(l)}(\{\beta J_0^{(\Sigma;l,k)}\};0)$ start to be
non zero and to have alternated directions in some ordered way 
to give rise to a pure antiferromagnetism so that, from Eqs. (\ref{App7}) and (\ref{App11}), one could have 
in principle $m^{(\mathrm{F})}=0$ but $\{m^{(\mathrm{F};l)}\}\neq\{0\}$.
On the other hand, at sufficiently low temperatures the SG solution 
(whose effective short- and long-range couplings are all non negative) 
will become the leading solution. In fact, as an argument based on frustration suggests,
even if we are not able to give here the general proof, 
we expect that the pure antiferromagnetism of the pure model
(to which would correspond a zero order parameter $m^{(\mathrm{F})}$) is never able to win against the spin glass solution. 
In other words, as soon as $c\neq 0$ (and then $J^{(\mathrm{F})}\neq 0$), 
%if there is no asymmetry among the communities, or if there is no differentation among
%the distributions of the long-range coulings (as having for instance $d\mu^{(l,l)}\neq d\mu^{(l,k)}$ for $l\neq k$), 
in the homogeneous case,
for $J_0^{(1,2)}\geq0$ there is no way to have any kind of antiferromagnetism and, more in general,
even if $J_0^{(1,2)}<0$, antiferromagnetism - if any - is expected to be very weak and to be dumped
by the spin glass phases
(note however that for $c$ exactly zero a regular antiferromagnetism may set in if $J_0^{(1,2)}<0$). 
This result is quite natural
and, for $n=2$, in two dimensions, has been numerically confirmed in \cite{Herrero2}
with the choice $J^{(1,1)}=J^{(2,2)}=J^{(1,2)}=J_0^{(1,1)}=J_0^{(2,2)}=J_0^{(1,2)}<0$~\footnote{Note that the 
small-world model considered in \cite{Herrero2} is built by rewiring the bonds, rather than by
adding a random connectivity, but for small probabilities $p$ of rewiring or by adding 
random bonds with small connectivity $c$,
the two versions of the models are expected to give nearly same results by identifying
$p$ with $c$.}. 
We point out that Eqs. (\ref{App7})-(\ref{App11}) hold for any
choice of the parameters. In particular, they hold also for $n=N$
which amounts formally to a single community
(the result reported in the Appendix of the Ref. \cite{SW} referred to this choice).
%As a consequence, for the considerations we have just
%done above, we see that, for any $c> 0$, internal antiferromagnetism in the homogeneous case 
%cannot exist, or more precisely, is expected to be dumped by the spin glass phase.

In conclusion, the homogeneous case does not have antiferromagnetism: to have antiferromagnetism in small-world system,
it is not enough to have more communities but it is necessary that 
be present some differentiation among the distributions of the couplings or  
some asymmetry, either in the size, in the in- and out-couplings, or in
the external fields. Without any differentiation or asymmetry the whole collection of the communities
can stay only in the same ferromagnetic or spin glass phase without any long-range heterogeneity.
For example, a typical minimal condition to have some antiferromagnetism consists 
in taking, for any $l$, $J^{(l,l)}=0$ and, for any couple $(l,k)$, with $l\neq k$, all the $J^{(l,k)}$ 
distributed according to a distribution $d\mu^{(1,2)}$ having a negative average.
We will analyze this case, \textit{the symmetric case}, in detail in the next paragraph.

\subsubsection{The symmetric case - mutual antiferromagnetism}
The simplest non trivial case to see antiferromagnetism 
consists in choosing the parameters in such a way that we have
the same short-range coupling as well as the same effective long-range coupling inside any community, 
and another same effective long-range coupling between any two different communities.
This requires that, for any $l$, $c^{(l,l)}=c^{(1,1)}$,  $J_0^{(l)}=J_0$, $d\mu^{(l,l)}=d\mu^{(1,1)}$,
and, for any couple $(l,k)$ with $l\neq k$, $c^{(l,k)}=c^{(1,2)}$, $J_0^{(l,k)}=0$, and $d\mu^{(l,k)}=d\mu^{(1,2)}$, with $d\mu^{(1,2)}\neq d\mu^{(1,1)}$
Hence, for $\Sigma=$F or $\Sigma=$SG, we are left with the only three effective couplings: 
$\beta J_0^{(\Sigma;l,l)}=\beta J_0^{(\Sigma)}$, $\beta J^{(\Sigma;l,l)}=\beta J^{(\Sigma;1,1)}$ and,
for $l\neq k$, $\beta J^{(\Sigma;l,k)}=\beta J^{(\Sigma;1,2)}\neq \beta J^{(\Sigma;1,1)}$.
Note that the condition $c^{(l,k)}=c^{(1,2)}$, for $l\neq k$, requires the equalities of the
relative sizes $\alpha^{(l)}=1/n$.
In this case (the symmetric case) the matrix $A^{(\Sigma;l,k)}$ simplifies in
\begin{eqnarray}
\label{AF}
A^{(\Sigma;l,k)}&=& b\delta_{l,k}-x\left(1-\delta_{l,k}\right),
\end{eqnarray}
%\begin{eqnarray}
%\label{AF}
%A^{(\Sigma;l,k)}&=&\delta_{l,k}\left(1-\beta J^{(\Sigma;1,1)}\right)-
%\beta J^{(\Sigma)}\tilde{\chi}_0\left(\beta J_0^{(\Sigma)};0\right)\left(1-\delta_{l,k}\right),
%\end{eqnarray}
where 
\begin{eqnarray}
\label{AF1}
b\defi 1-\beta J^{(\Sigma;1,1)}\tilde{\chi}_0\left(\beta J_0^{(\Sigma)};0\right),
\end{eqnarray}
and
\begin{eqnarray}
\label{AF2}
x\defi \beta J^{(\Sigma;1,2)}\tilde{\chi}_0\left(\beta J_0^{(\Sigma)};0\right).
\end{eqnarray}
Hence, in the symmetric case the determinant can be explicitly calculated as
\begin{eqnarray}
\label{AF3}
\det \bm{A}^{(\Sigma)}&=&\left(b+x\right)^{n-1}\left(b-x\left(n-1\right)\right).
\end{eqnarray}
%\begin{eqnarray}
%\label{AF3}
%\det \bm{A}^{(\Sigma)}&=&\left(1-\beta J^{(\Sigma)}\tilde{\chi}_0\left(\beta J_0^{(\Sigma)};0\right)\right)^{n-1}
%\nonumber \\
%&\times& \left(1+\beta J^{(\Sigma)}\tilde{\chi}_0\left(\beta J_0^{(\Sigma)};0\right)\left(n-1\right)\right).
%\end{eqnarray}
From Eq. (\ref{AF3}) we see that Eq. (\ref{THEOs2}) for the critical temperature 
of a second-order phase transition has two solutions: $x=-b$ and $x=b/(n-1)$. 
%The first solution occurs when the two measures $d\mu^{(1,1)}$ and $d\mu^{(1,2)}$ 
%have averages of same sign, while the
%second when their averages have opposite signs. 
Therefore, we have the two following possible solutions
\begin{eqnarray}
\label{AF4}
&&\left(\beta_c^{(\Sigma)} J^{(\Sigma;1,2)}-\beta_c^{(\Sigma)} J^{(\Sigma;1,1)}\right)
\nonumber \\
&\times& \tilde{\chi}_0\left(\beta_c^{(\Sigma)} J_0^{(\Sigma)};0\right)=-1,
\end{eqnarray}
\begin{eqnarray}
\label{AF5}
&&\left(\left(n-1\right)\beta_c^{(\Sigma)} J^{(\Sigma;1,2)}+\beta_c^{(\Sigma)} J^{(\Sigma;1,1)}\right)
\nonumber \\
&\times&\tilde{\chi}_0\left(\beta_c^{(\Sigma)} J_0^{(\Sigma)};0\right)=1.
\end{eqnarray}
For $\Sigma=$SG Eq. (\ref{AF5}) gives always a solution, whereas for $\Sigma=$F a possible
solution will come either from Eq. (\ref{AF4}) or from Eq. (\ref{AF5}) 
according to the signs of the effective couplings $J^{(\mathrm{F};1,1)}$ and $J^{(\mathrm{F};1,2)}$, which are  
averages with respect to the given measures $d\mu^{(1,1)}$ and $d\mu^{(1,2)}$, respectively. 
If we are sufficiently far from the homogeneous case $\beta J^{(1,1)}= \beta J^{(1,2)}$,
antiferromagnetism can take place at a temperature given by Eq. (\ref{AF4}).  
We can distinguish in turn the symmetric case in two sub-cases.\\

\textit{Only mutual interaction:} $\beta J^{(\mathrm{F};1,1)}=0$.
If $d\mu^{(1,1)}$ has zero average and $\beta J^{(\mathrm{F};1,2)}<0$, the solution for
$\beta_c^{(\mathrm{F})}$ comes only from Eq. (\ref{AF4}).  
In this last case it is easy 
to see that antiferromagnetism sets in by observing that
for the self-consistent system (\ref{THEO})-(\ref{THEO1}) 
there are for instance always solutions of the form  
\begin{eqnarray}
\label{AF6}
&&\left(m^{(\mathrm{F};1)},\ldots,m^{(\mathrm{F};n)}\right)=\nonumber \\
&&\left(0,\ldots,0,m^{(\mathrm{F})},0,\ldots,0,-m^{(\mathrm{F})},0,\ldots,0\right),
\end{eqnarray}
and all its combinations, where $m^{(\mathrm{F})}$ is any solution of
\begin{eqnarray}
\label{AF7}
&& m^{(\mathrm{F})}=\nonumber \\
&& m_0\left(\beta J_0^{(\mathrm{F})};\beta J^{(\mathrm{F};1,2)}
m_0\left(\beta J_0^{(\mathrm{F})};\beta J^{(\mathrm{F};1,2)} m^{(\mathrm{F})}\right)\right).
\end{eqnarray}
Note that, due to the parity of the function $m_0$ with respect to its second argument,
in Eq. (\ref{AF7}) we are free to substitute $J^{(\mathrm{F};1,2)}$ with $|J^{(\mathrm{F};1,2)}|$.
Solutions as Eqs. (\ref{AF6})-(\ref{AF7}) 
are evidently antiferromagnetic and for sufficiently high temperatures are leading
against the SG solution. In Fig. 1 we plot a case with $J^{(1,1)}=0$ and $J^{(1,2)}=1.5$.
Observe that in this case the solution $m^{(\mathrm{F})}\neq 0$ is never stable under iteration. 
Finally note that, for $J_0<0$ and $\Sigma=$F, the two terms appearing in the 
lhs of Eqs. (\ref{AF4}) and (\ref{AF5}) are no more monotonic functions of $\beta$,
so that, for a sufficiently large value of $c^{(1,2)}$, 
$\beta_c^{(\mathrm{F})}$ will have multiple solutions. Furthermore, by observing that
all the critical behavior of the system is encoded in the single susceptibility
$\tilde{\chi}_0(\beta J_0^{(\Sigma)};0)$, we see that we can use the same analysis performed 
in the Ref. \cite{SW}: when $J_0<0$ the non monotonicity reflects also on the
fact that the P-F transition may be of first-order.\\

\textit{Mutual and internal interaction:} $\beta J^{(1,2)},\beta J^{(1,1)}\neq 0$.
Much more interesting is the case in which there are also internal long range-couplings.
Let us consider for example the case with two communities. The self-consistent
system (\ref{THEO}) reduces to
\begin{widetext}
\begin{eqnarray}
\label{AF11}
m^{(\Sigma;1)}&=&m_0^{(1)}\left(\beta J_0^{(1)};\beta J^{(\Sigma;1,1)}m^{(\Sigma;1)}+\beta J^{(\Sigma;1,2)}m^{(\Sigma;2)}\right),
\nonumber \\
m^{(\Sigma;2)}&=&m_0^{(2)}\left(\beta J_0^{(2)};\beta J^{(\Sigma;1,2)}m^{(\Sigma;1)}+\beta J^{(\Sigma;1,1)}m^{(\Sigma;2)}\right),
\end{eqnarray}
\end{widetext}
where we have used the fact that in the symmetric case $\beta J^{(\Sigma;2,2)}=\beta J^{(\Sigma;1,1)}$ and
$\beta J^{(\Sigma;2,1)}=\beta J^{(\Sigma;1,2)}$. 
In the previous case with $\beta J^{(1,1)}=0$ essentially we had just one order parameter 
given by Eq. (\ref{AF7}), in the present case instead such a reduction is not possible. 
Here there are two order parameters which are intrinsically not zero due to the presence of the
internal coupling $\beta J^{(1,1)}$ and at the same time the two order parameters interact
through the coupling $\beta J^{(1,2)}$ whose sign determines whether they are parallel or anti-parallel. 
For $n=2$ from Eqs. (\ref{AF4}) and (\ref{AF5}) we see that, if $\beta J^{(1,1)}>0$, we have two critical temperatures if 
$\beta J^{(1,1)}$ is sufficiently bigger than $|\beta J^{(1,2)}|$ (and similarly if $\beta J^{(1,1)}<0$),
so that more interesting phenomena are expected in this case.
The general mechanism will result clearer in the next paragraph.

\begin{figure}
\epsfxsize=65mm \centerline{\epsffile{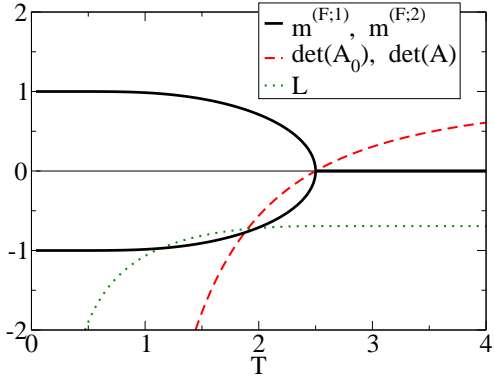}}
\caption{(Color on-line) Solution of the self consistent system (\ref{THEO}) for the CW model in the symmetric 
case $n=2$ with
$J^{(1,1)}=0$ and $J^{(1,2)}=2.5$ or $-J^{(1,2)}=2.5$. There is only one critical temperature located at $T_c=2.5$. 
Note that here we plot all the possible solutions. For $J^{(1,2)}>0$, $m^{(\mathrm{F};1)}$ and $m^{(\mathrm{F};2)}$
(continuous lines) are parallel and coincide, whereas for $J^{(1,2)}<0$ they are anti-parallel. 
The plot of $\det(A)$ (dashed line) represents the stability curve - under iteration - 
of the solution $(m^{(\mathrm{F};1)},m^{(\mathrm{F};2)})$, Eq. (\ref{THEOs}),
and similarly the plot of $\det(A_0)$ represents the stability curve 
for the trivial solution $m^{(\mathrm{F};1)}=m^{(\mathrm{F};2)}=0$;
in the case of only mutual interactions the two coincide: $\det(A)=\det(A_0)$. 
Clearly, in this example, for $T<T_c$, no solution results stable under iteration.
We plot also the free energy term $L$ (dotted line) applied, via Eq. (\ref{THEO12}), 
to the non trivial solution $(m^{(\mathrm{F};1)},m^{(\mathrm{F};2)})$. The free energy term $L$ 
obtained applied instead to the trivial solution 
$m^{(\mathrm{F};1)}=m^{(\mathrm{F};2)}=0$ is a constant (for more details about 
the CW model see Sec. VII), $L=L_0=-\log(2)$ and is plotted only in the region
where the trivial solution is the unique (and - of course - stable) solution.} 
\label{fig1}
\end{figure}

\subsubsection{The general case}
Of course making an analysis of the general case is a formidable task,
however we can get important insights by looking first at the simplest CW model.
In the CW model we have two great simplifications: since there is no short-range coupling
we have $\tilde{\chi}_0\equiv 1$, furthermore, since in the CW model the connectivities 
become infinite in the thermodynamic limit, 
the effective couplings $\beta J^{(\Sigma;l,k)}$ become linear in $\beta$ (see later for details).
As a consequence, for the CW model in the symmetric case Eqs. (\ref{AF4}) and (\ref{AF5}) are linear in $\beta$ and
may have at most one solution each. However, as soon as we are not in the symmetric case, the general Eq. (\ref{THEOs2})
determining a critical temperature of a second-order phase transition 
is no longer linear and can have many solutions, \textit{i.e.}, when we are not in the symmetric
the degeneracy featuring Eq. (\ref{THEOs2}) disappears and
we may have a number of multicritical points $\mathop{O}(n)$ where one or more order parameters are non analytic.
As in the case of a single community, $n=1$, the
existence of multicritical points gives us a clue about the fact that the self-consistent 
equations admit more stable solutions. 
%and we may also have first-order
%phase transitions with finite jumps of the order parameters. 
In Fig. 2 we plot a case with $n=2$. Note however that in the case $n=1$
the necessary condition for the existence of multicritical points was to have a short-range 
coupling negative, while now we do not require this condition, the mechanism is indeed completely
different and based on the fact that there is some differentiation among the couplings.
In \cite{Contucci} was shown that, for sufficiently large $J$'s, 
the CW model (see later Sec. V) with $n=2$ exhibits a first-order phase transition 
tuned by the relative sizes of the two communities. In our approach this fact is clear since
the effective couplings $J^{(\Sigma;l,k)}$, given by Eqs. (\ref{THEO1}) and (\ref{THEO2}), 
depend on the connectivities $c^{(l,k)}$
which in turn depend on the relative sizes $\alpha^{(l)}$ through Eqs. (\ref{conn1}); as soon as the relative sizes are not
equal the connectivities $c^{(l,k)}$ and then the effective couplings $J^{(\Sigma;l,k)}$ are no more symmetric
so that more stable solutions may exist, and, by varying the parameters and keeping fixed the temperature,
we pass from one solution, say $\{m^{(\Sigma;l)}\}$, to the other, say $\{m^{'(\Sigma;l)}\}$, 
performing finite jumps, \textit{i.e.}, first-order phase transitions in the space
of the free parameters, with probability 1.

If we consider now a bit more complicated model as the VB model, we still have $\tilde{\chi}_0\equiv 1$
since there is no short-range coupling, but now the connectivities are finite so that
the effective couplings are no more linear in $\beta$. However, in general, for $n$ not small,
hardly the symmetric case with only positive couplings will give multicritical points through 
Eqs. (\ref{AF4}) and (\ref{AF5})
(this can be understood considering large but finite connectivities);
to have multicritical points, and, in the space of the parameters, possible first-order phase transitions, 
it will be necessary to be far from the symmetric case with
some differentiation among the effective couplings. 
Finally, we expect that also in the case of positive short-range couplings,
this scenario basically holds as well. 
However, when some of the short-range couplings are negative, the scenario of course changes completely
and, as we already know from the case $n=1$, we may have multicritical points and 
first-order phase transitions also with respect to the temperature and 
even in the symmetric case, \textit{i.e.}, 
without the necessity to have some differentiation among the effective couplings.

Recall that, in general, only one solution of the self-consistent equations is leading in the thermodynamic limit
and, furthermore, a phase transition itself may be not leading in this limit,
however the stable not leading solutions may play an important role when $n$ is high (see next paragraph).

\begin{figure}
\epsfxsize=65mm \centerline{\epsffile{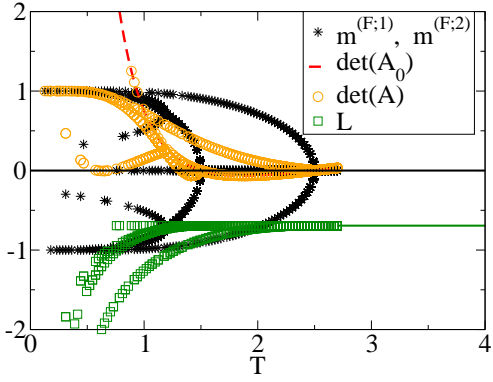}}
\caption{(Color online) Solution of the self consistent system (\ref{THEO}) 
for the CW model in the symmetric case $n=2$ with
$J^{(1,1)}=2$ and $J^{(1,2)}=2.5$, or $J^{(1,2)}=-2.5$. 
Note that here we plot all the possible solutions. For $J^{(1,2)}>0$, $m^{(\mathrm{F};1)}$ and $m^{(\mathrm{F};2)}$
(stars) are parallel, whereas for $J^{(1,2)}<0$ they are anti-parallel.
The several branches of $\det(A)$ (circles) represent the stability - under iteration - 
of the several non trivial solutions $(m^{(\mathrm{F};1)},m^{(\mathrm{F};2)})$, Eq. (\ref{THEOs}),
and similarly the plot of $\det(A_0)$ (dashed line) represents the stability for the trivial solution 
$m^{(\mathrm{F};1)}=m^{(\mathrm{F};2)}=0$.
We plot also the several branch's of the free energy term $L$ (squares) applied, via Eq. (\ref{THEO12}), to the several solutions. 
Here Eqs. (\ref{AF4}) and (\ref{AF5}) give two instability points
at the temperatures $T_{c1}=2.5$ and $T_{c2}=1.5$. 
In the thermodynamic limit, only  $T_{c1}$ corresponds to a true critical temperature, whereas the other
corresponds to a metastable solution. Furthermore we see another metastable solution at $T_{c3}=1.18$
featured as two broken symmetries
where $(m^{(\mathrm{F};1)},m^{(\mathrm{F};2)})$ do not transit around 0, but around the values $\pm$0.67.} 
\label{fig2}
\end{figure}

\begin{figure}
\epsfxsize=65mm \centerline{\epsffile{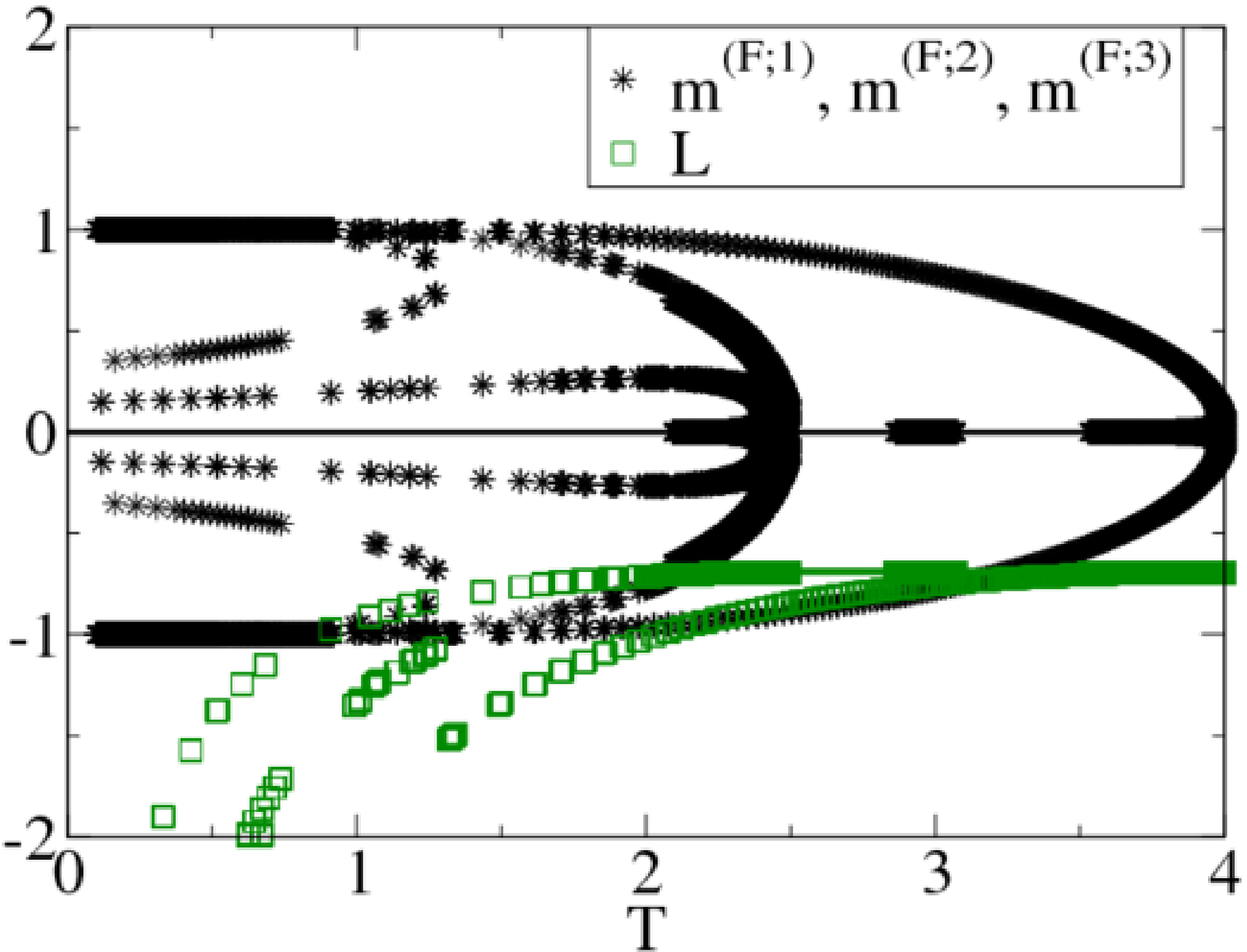}}
\caption{(Color online) Solution of the self consistent system (\ref{THEO}) 
for the CW model in the symmetric case $n=3$ with
$J^{(1,1)}=3$ and $J^{(1,2)}=0.5$. 
Note that here we plot all the possible solutions.
We plot also the several branches of the free energy term $L$ applied, via Eq. (\ref{THEO12}), to the several solutions. 
Here Eqs. (\ref{AF4}) and (\ref{AF5}) give two instability points
at the temperatures $T_{c1}=4$ and $T_{c2}=2.5$. 
In the thermodynamic limit, only  $T_{c1}$ corresponds to a true critical temperature, whereas the other
corresponds to a metastable solution. Furthermore we see another metastable solution at $T_{c3}=1.32$
featured as two broken symmetries
where $(m^{(\mathrm{F};1)},m^{(\mathrm{F};2)},m^{(\mathrm{F};3)})$ 
do not transit around 0, but around the values $\pm$0.75.} 
\label{fig3}
\end{figure}

\begin{figure}
\epsfxsize=65mm \centerline{\epsffile{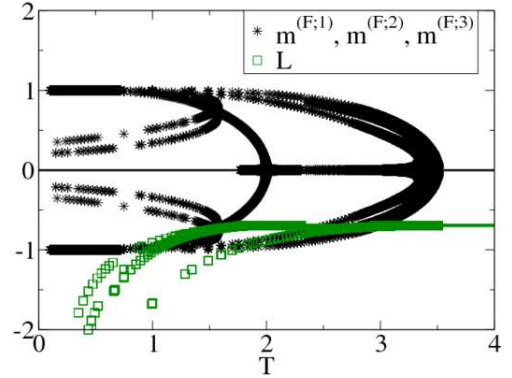}}
\caption{(Color online) Solution of the self consistent system (\ref{THEO}) 
for the CW model in the symmetric case $n=3$ with
$J^{(1,1)}=3$ and $J^{(1,2)}=-0.5$. 
Note that here we plot all the possible solutions. 
We plot also the several branches of the free energy term $L$ applied, via Eq. (\ref{THEO12}), to the several solutions. 
Here Eqs. (\ref{AF4}) and (\ref{AF5}) give two instability points
at the temperatures $T_{c1}=3.5$ and $T_{c2}=2$. 
In the thermodynamic limit, only  $T_{c1}$ corresponds to a true critical temperature, whereas the other
corresponds to a metastable solution. We see further metastable solutions at $T_{c3}=1.6$.
with multiple broken symmetries
where $(m^{(\mathrm{F};1)},m^{(\mathrm{F};2)},m^{(\mathrm{F};3)})$ 
do not transit around 0, but around the values $\pm$0.75 and 
$\pm$0.80.} 
\label{fig4}
\end{figure}

\subsubsection{Behavior for large $n$}
As it was already evident from the previous paragraphs, if there is some differentiation among
the communities, as $n$ increases the number of solutions of the self-consistent system grows.
In Fig. \ref{fig3} we report a case with n=3.
Now, only one of these solutions is leading, nevertheless, the other stable not leading solutions
as metastable states play a more and more important role in the limit of large $n$, especially 
when some of the communities interact through negative couplings.
In Fig. \ref{fig4} we report a case with n=3 and negative inter-couplings.
Indeed, coming back to the symmetric case, from Eqs. (\ref{AF4}) and (\ref{AF5}) we see that when the number of
communities is large, $n\gg 1$, and the communities are connected, 
the highest critical temperature comes only from Eq. (\ref{AF5}) and is the solution of the following equation
\begin{eqnarray}
\label{AF12}
\beta_c^{(\Sigma)} J^{(\Sigma;1,2)}\tilde{\chi}_0\left(\beta_c^{(\Sigma)} J_0^{(\Sigma)};0\right)\simeq\frac{1}{n}.
\end{eqnarray}
Recalling the definition of $J^{(\Sigma;1,2)}$ we see that the transition described by Eq. (\ref{AF12}) 
will be either P-F or P-SG. In particular, if $J^{(\mathrm{F};1,2)}\leq 0$ the leading transition
will be only P-SG.

In more realistic situations, the system will be far from the symmetric case.
Typically, different couples of communities will be coupled by different couplings
of arbitrary amplitudes and signs, implying therefore frustration.
As a consequence, in such a disordered structure the expected leading transitions will be P-SG.
On the other hand such a claim results immediately clear to the reader familiar
with spin-glass theory. In fact, if we look at our self-consistent equations
(\ref{THEO})-(\ref{THEO1}) (or (\ref{THEOG})), for example considering the simplest CW
case, apart from the fact that in these equations there is not the Onsager's reaction term \cite{Onsager},
they are formally identical to the TAP equations \cite{TAP} 
(recall that $\beta H^{(\Sigma;l)}$ is the field seen by $l$-th community 
due to the self-magnetization $m^{(\Sigma;l)}$ and to the others $n-1$ magnetizations). 
Yet, due to our general mapping which establishes a strong universality of all the Poisson small-world models, 
the general structure of these equations roughly speaking
is still of the TAP kind - but without the Onsager's term - even when there are short-range couplings.
Therefore, the typical multi-valley landscape scenario representing the existence of infinitely many
metastable states (whose number grows exponentially with $n$), 
possibly also separated by infinitely high free energy barriers, is expected in
the limit of large $n$ with negative long range couplings \cite{Parisi},\cite{Fisher}.
We recall that at low temperatures, where the glassy scenario has been exactly confirmed in the SK case, 
the Onsager's term disappears \cite{Parisi}. 
%We recall that in the CW case, the TAP equations without the Onsager's term have been studied
%in \cite{Gross} for the $p$-spin model with $p\to\infty$ and, at $T=0$, the glassy scenario
%was exactly confirmed.

\subsubsection{Attaching good to bad communities}
When in a given community the $J$'s are almost absent and the $J_0$'s are negative,
or either when the $J$'s are negative (at least in average),
as we have already learned, there is no way
for such a bad community to have any long-range order, and the only possible state, at low temperature, 
inside this bad community,
kept isolated, is the glassy state (provided that its own connectivity be sufficiently high).
An interesting issue is then to understand what happens if we connect 
the bad community with a good community (\textit{i.e.}, having positive interactions and then some long-range order) 
through a certain number (proportional to some added connectivity $c{(1,2)}$) of random couplings $J^{(1,2)}$.
The perhaps surprising answer is that, not only the bad community gains an order,
but also the already good community improves its order.
We point out that this result is not immediately so obvious a priori; the fact that the
number of interactions per spin has been increased by $c{(1,2)}$ cannot be used 
to explain this effect that takes place for both positive or negative  
random couplings $J^{(1,2)}$.
However, a simple argument based on the high temperature expansion shows actually that
this is the case: increasing the average connectivity always improve the order.
In Fig. 5 we report an example for the VB model and compare the two situations
with and without attaching the two communities.

\begin{figure}
\epsfxsize=65mm \centerline{\epsffile{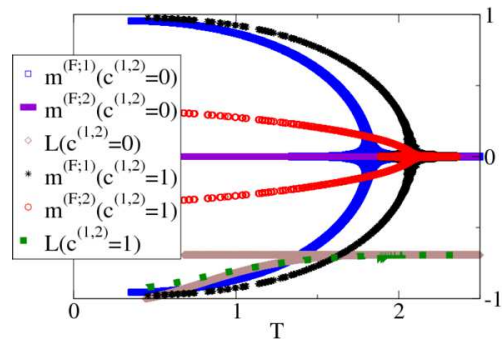}}
\caption{(Color online) Solutions of the self consistent system (\ref{THEO}) 
for the VB model $n=2$ with couplings
$J^{(1,1)}=1$, $J^{(2,2)}=-1$ and $J^{(1,2)}=-1$ or $J^{(1,2)}=1$,
and connectivities $c^{(1,1)}=2$, $c^{(2,2)}=2$, and $c^{(1,2)}=0$ 
(empty squares for $m^{(\mathrm{F};1)}$ and continuous line for $m^{(\mathrm{F};2)}$) 
or $c^{(1,2)}=1$ (stars for $m^{(\mathrm{F};1)}$ and circles for $m^{(\mathrm{F};2)}$).  
We plot also the free energy terms $L$ (rhombus and filled squares for the two cases).} 
\label{fig5}
\end{figure}

\subsection{Level of accuracy of the method}
%The self-consistent equations (\ref{THEOa}) and the correlation functions (\ref{THEOh})
As anticipated in the introduction, the level of accuracy of this effective field theory
is the same as that discussed in the Ref. \cite{SW}. More precisely,   
Eqs. (\ref{THEO}-\ref{THEO12G}) are exact in the P region,
\textit{i.e.}, the region where any of the $2n$ order parameters is zero;
whereas in the other regions
provide an effective approximation whose level of accuracy depends  
on the details of the model. In particular, in the absence of frustration
the method becomes exact at any temperature in two important limits:
in the limit $c^{(l,k)}\to 0^+$, $l,k=1,\ldots,n$, 
in the case of second-order phase transitions, due
to a simple continuity argument;
and in the limit $c^{(l,k)}\to\infty$, $l,k=1,\ldots,n$, due to the fact that in this
case the system becomes a suitable fully connected model exactly described
by the self-consistent equations (\ref{THEO}) (of course, 
when $c^{(l,k)}\to\infty$, to have a finite critical temperature one has to renormalize 
the average of the shortcuts couplings by $c^{(l,k)}$).  

\section{Examples}
In this section we consider some applications for which the method can
be fully applied analytically: the generalized VB model and the generalized 
one-dimensional small-world model. By using our effective field theory, both these models have been already studied in
the Ref. \cite{SW} in their $n=1$ version.  
Within the VB models, we will consider in particular the limit in which all connectivities
go to infinity, obtaining then the generalized CW and the generalized SK models.
The CW case model can be extrapolated also from the SK model as a particular
(actually the simplest) case.

\begin{widetext}
\subsection{Generalized Viana Bray models and the generalized Curie-Weiss and Sherrington-Kirkpatrick limits}
Among all the possible cases of the models introduced in Sec. II, the simplest family of models  
is the one with no short-range couplings: $J_0^{(l,k)}=0$, for any $l,k$,
which, can be seen as a generalization of the Viana-Bray model to $n>1$ communities. 
When there is no short-range coupling the pure magnetizations and the 
corresponding free energy densities of each community read as
\begin{eqnarray}
\label{VB0}
m_0^{(l)}\left(0;\beta h^{(l)}\right)=\tanh\left(\beta h^{(l)}\right), 
\end{eqnarray}
\begin{eqnarray}
\label{VB}
\beta f_0^{(l)}\left(0;\beta h^{(l)}\right)=-\log\left[2\cosh\left(\beta h^{(l)}\right)\right], 
\end{eqnarray}
so that the system (\ref{THEO})-(\ref{THEO1}) becomes
\begin{eqnarray}
\label{VB1}
m^{(\Sigma;l)}=\tanh\left(H^{(\Sigma;l)}+\beta h^{(l)}\right), 
\end{eqnarray}
\begin{eqnarray}
\label{VB1h}
H^{(\Sigma;l)}=\sum_k \beta J^{(\Sigma;l,k)}m^{(\Sigma;k)},
\end{eqnarray}
and for the Landau free energy density one has
\begin{eqnarray}
\label{VB1f}
L^{(\Sigma)}\left(m^{(\Sigma;1)},\ldots,m^{(\Sigma;n)}\right)&=&\sum_l\alpha^{(l)}
\left\{\frac{\beta J^{(\Sigma;l,l)}\left(m^{(\Sigma;l)}\right)^2}{2}
 -\log\left[2\cosh\left(\beta H^{(\Sigma;l)}\right)\right]\right\}
\nonumber \\
&+& \sum_{l\neq k} \frac{\alpha^{(l)}\beta J^{(\Sigma;l,k)}m^{(\Sigma;l)}m^{(\Sigma;k)}}{2},
\end{eqnarray} 
where the effective couplings $\beta J^{(\Sigma;l,k)}$ are given by Eqs. (\ref{THEO2}) and (\ref{THEO3}).
%and it is understood that $H^{(\Sigma;l)}$
%in Eq. (\ref{VB1f}) is function of the free variables $\{m^{(l)}\}$, not of the solutions $\{m^{(\Sigma;l)}\}$.
In particular, more explicitly, for the measure (\ref{mea}) we have %the system (\ref{VB1}) reads as
\begin{eqnarray}
\label{VB2}
m^{(\Sigma;l)}=\tanh\left(\sum_k \tanh^{l_\Sigma}\left(\beta J^{(l,k)}\right)c^{(l,k)}m^{(\Sigma;k)}+\beta h^{(l)}\right), 
\end{eqnarray}
where $l_\Sigma=$1 or 2 for $\Sigma=$F or SG, respectively; whereas for
the measure (\ref{mea1}) one has %the system (\ref{VB1}) becomes
\begin{eqnarray}
\label{VB3}
m^{(\mathrm{F};l)}=\tanh\left(\beta h^{(l)}\right), 
\end{eqnarray}
and 
\begin{eqnarray}
\label{VB4}
m^{(\mathrm{SG};l)}=\tanh\left(\sum_k \tanh^{2}\left(\beta J^{(l,k)}\right)c^{(l,k)}m^{(\mathrm{SG};k)}+\beta h^{(l)}\right), 
\end{eqnarray}
so that in the case of the measure (\ref{mea1}) only P-SG like phase transitions are possible.
In the case of the measure (\ref{mea}) from Eq. (\ref{VB2}) we see that if all the couplings $J^{(l,k)}$
are non negative, there are both P-F and P-SG like phase transitions, but only the former
are leading, due to their higher critical temperature. However, if some of the couplings $J^{(l,k)}$
is negative, in general there can be P-F like phase transitions with competitions between 
ferromagnetism and antiferromagnetism and, furthermore, for some range of the parameters there can be also
a competition with a P-SG like phase transition that in turn gives rise to $n$ stable spin glass like order
parameters $m^{(\mathrm{SG};l)}$.

The inverse critical temperature $\beta_c^{(\Sigma)}$
of any possible second-order phase transition can be obtained by developing
the self-consistent system (\ref{VB1})-(\ref{VB1h}) for small $m^{(\Sigma;l)}$ and $h^{(l)}=0$. 
As we have seen in Sec. IIIB, this amounts to find the solutions of  
the equation $\det (A^{(\Sigma;l,k)})=0$ 
which in the present case becomes 
\begin{eqnarray}
\label{VB5}
A^{(\Sigma;l,k)}&=&\delta_{l,k} -\beta_c^{(\Sigma)} J^{(\Sigma;l,k)},
\end{eqnarray}
where we have used $\tilde{\chi}_0^{(l,l)}(0;\beta h^{(l)})=1-\tanh^2(\beta h^{(l)})$.
%for $\{J_0^{(l,k)}\}=\{0\}$ and $\{h^{(l)}\}=\{0\}$, we see that
%the solution $m^{(\Sigma;l)}\equiv 0$ becomes unstable when $\det (A^{(\Sigma;l,k)})=0$
The non linear equation $\det \bm{A}^{(\Sigma)}=0$ provides the exact critical temperature 
of any second-order phase transition of this generalized
Viana-Bray model. For example, for $n=2$ this equation becomes:
\begin{eqnarray}
\label{VB6}
\left(1-\beta J^{(\Sigma;1,1)}\right)\left(1-\beta J^{(\Sigma;2,2)}\right)-
\beta J^{(\Sigma;1,2)}\beta J^{(\Sigma;2,1)}=0,
\end{eqnarray}
which, for the measure (\ref{mea}), amounts to
%\begin{eqnarray}
%\label{VB7}
%&1&-\left(c^{(1,1)}\tanh^{l_\Sigma}\left(\beta J^{(1,1)}\right)+
%c^{(2,2)}\tanh^{l_\Sigma}\left(\beta J^{(2,2)}\right)\right)
%+c^{(1,1)}c^{(2,2)}\tanh^{l_\Sigma}\left(\beta J^{(1,1)}\right)\tanh^{l_\Sigma}\left(\beta J^{(2,2)}\right) 
%\nonumber \\
%&-&c^{(1,2)}c^{(2,1)}\left(\tanh^{l_\Sigma}\left(\beta J^{(1,2)}\right)\right)^2=0.
%\end{eqnarray}
\begin{eqnarray}
\label{VB8}
&1&-\left(c^{(1,1)}\tanh^{l_\Sigma}\left(\beta J^{(1,1)}\right)+c^{(2,2)}\tanh^{l_\Sigma}\left(\beta J^{(2,2)}\right)\right)
+c^{(1,1)}c^{(2,2)}\tanh^{l_\Sigma}\left(\beta J^{(1,1)}\right)\tanh^{l_\Sigma}\left(\beta J^{(2,2)}\right)
\nonumber \\
&-&\frac{\alpha}{1-\alpha}\left(c^{(1,2)}\tanh^{l_\Sigma}\left(\beta J^{(1,2)}\right)\right)^2=0,
\end{eqnarray}
where we have used $\alpha\defi\alpha^{(1)}$, $\alpha^{(2)}=1-\alpha$ and Eq. (\ref{conn1}).
As we have seen in Sec. IIIB, the symmetric case can be explicitly worked out also for
$n$ generic. From Eq. (\ref{AF4}) and (\ref{AF5}) we have
\begin{eqnarray}
\label{VB9}
\beta_c^{(\Sigma)} J^{(\Sigma;1,2)}-\beta_c^{(\Sigma)} J^{(\Sigma;1,1)}=-1,
\end{eqnarray}
and
\begin{eqnarray}
\label{VB10}
\left(n-1\right)\beta_c^{(\Sigma)} J^{(\Sigma;1,2)}+\beta_c^{(\Sigma)} J^{(\Sigma;1,1)}=1.
\end{eqnarray}

\subsubsection{The CW and the SK limits}
Let us now consider the Curie-Weiss limit with $\Sigma$=F. 
Given the set of the relative sizes $\{\alpha^{(l)}\}$, we can recover the Curie-Weiss limit 
by choosing the connectivities $c^{(l,k)}$ to be equal to their maximum value
which, according to Eq. (\ref{conn3}), is given by
\begin{eqnarray}
\label{CW}
c^{(l,k)}=\alpha^{(k)}N.
\end{eqnarray}
With this choice the probabilities $p^{(l,k)}$ become
\begin{eqnarray}
\label{CW1}
 p(c_{ij}^{(l,k)})&=&\delta_{c_{ij}^{(l,k)},1},
\end{eqnarray}
so that, by choosing the measure (\ref{mea}) with the $J^{(l,k)}$ renormalized by $N$
\begin{eqnarray}
\label{CW2}
\frac{d\mu^{(l,k)}}{dJ_{i,j}^{(l,k)}}=\delta\left(J_{i,j}^{(l,k)}-\frac{J^{(l,k)}}{N}\right),
\end{eqnarray}
the CW limit is recovered (of course anything can be rephrased 
in terms of limits by simply substituting the lhs of Eqs. (\ref{CW})-(\ref{CW2})
the equalities with arrows-limits for $N\to\infty$).
By using Eqs. (\ref{CW})-(\ref{CW2}) for large $N$, the system (\ref{VB1})-(\ref{VB1h}) and the 
Landau free energy (\ref{VB1f}), for $\Sigma=$F, become 
\begin{eqnarray}
\label{CW3}
m^{(\mathrm{F};l)}=\tanh\left(H^{(\mathrm{F};l)}+\beta h^{(l)}\right), 
\end{eqnarray}
\begin{eqnarray}
\label{CW4}
H^{(\mathrm{F};l)}=\sum_k \beta J^{(l,k)}\alpha^{(k)}m^{(\mathrm{F};k)},
\end{eqnarray}
\begin{eqnarray}
\label{CW5}
L^{(\mathrm{F})}\left(m^{(\Sigma;1)},\ldots,m^{(\Sigma;n)}\right)&=&\sum_l\alpha^{(l)}
\left\{\frac{\alpha^{(l)}\beta J^{(l,l)}\left(m^{(\Sigma;l)}\right)^2}{2}
 -\log\left[2\cosh\left(\beta H^{(\mathrm{F};l)}\right)\right]\right\}
\nonumber \\
&+& \sum_{l\neq k} \frac{\alpha^{(l)}m^{(\Sigma;l)}\beta J^{(l,k)}\alpha^{(k)}m^{(\Sigma;k)}}{2},
\end{eqnarray} 
which generalizes the result found in \cite{Contucci} for $n=2$ to general $n$.
%(it is understood that $H^{(\mathrm{F};l)}$
%in Eq. (\ref{CW5}) is function of the free variables $\{m^{(\Sigma;l)}\}$, not of the solutions $\{m^{(\mathrm{F};l)}\}$).
Notice that, as we explain in Sec. IIIC, Eqs. (\ref{CW3})-(\ref{CW5}) are exact at any temperature.

Analogously, if we assume again (\ref{CW}) and choose the measure (\ref{mea2}),
the SK limit is recovered, and for large $N$ the system (\ref{VB1})-(\ref{VB1h}) and the 
Landau free energy (\ref{VB1f}), become 
%(it is understood that $H^{(\Sigma;l)}$
%in Eq. (\ref{CW8}) is function of the free variables $\{m^{(l)}\}$, not of the solutions $\{m^{(\Sigma;l)}\}$)
\begin{eqnarray}
\label{CW6}
m^{(\Sigma;l)}=\tanh\left(H^{(\Sigma;l)}+\beta h^{(l)}\right), 
\end{eqnarray}
\begin{eqnarray}
\label{CW7}
H^{(\Sigma;l)}=\sum_k \beta J^{(\Sigma;l,k)}m^{(\Sigma;k)},
\end{eqnarray}
\begin{eqnarray}
\label{CW8}
L^{(\Sigma)}\left(m^{(1)},\ldots,m^{(n)}\right)&=&\sum_l\alpha^{(l)}
\left\{\frac{\beta J^{(\Sigma;l,l)}\left(m^{(l)}\right)^2}{2}
 -\log\left[2\cosh\left(\beta H^{(\Sigma;l)}\right)\right]\right\}
\nonumber \\
&+& \sum_{l\neq k} \frac{\alpha^{(l)}m^{(l)}\beta J^{(\Sigma;l,k)}m^{(k)}}{2},
\end{eqnarray} 
where for large $N$ the effective couplings are given by 
(according to the measure (\ref{mea2}) $J^{(l,k)}$ and $\tilde{J}^{(l,k)}$ are respectively 
the average and the variance of the couplings)
\begin{eqnarray}
\label{CW9}
\beta J^{(\mathrm{F};l,k)}= \alpha^{(k)} \beta J^{(l,k)},
\end{eqnarray}  
and
\begin{eqnarray}
\label{CW10}
\beta J^{(\mathrm{SG};l,k)}= \alpha^{(k)} \left(\beta \tilde{J}^{(l,k)}\right)^2.
\end{eqnarray}  
Eqs. (\ref{CW6})-(\ref{CW10}) generalize the result found in \cite{Almeida}
and \cite{MOI}, valid for only a uniform 
mutual coupling, to general mutual couplings as well as internal couplings. 
Notice that, as explained in Sec. IIIC, in unfrustrated systems, \textit{i.e.},
for $J^{(l,k)}\gg \tilde{J}^{(l,k)}$, Eqs. (\ref{CW6})-(\ref{CW10}) are exact at any temperature.

\subsection{One-dimensional small-world model for $n$ communities}
In the Ref. \cite{SW} we have studied in detail the one-dimensional small-world model
for $n=1$ emphasizing the existence of first-order phase transitions for 
negative short-range couplings. Here we generalize this result to $n$ communities.

For the free energy, the magnetization and the susceptibility of the 
pure model of the $l$-th community we have
\begin{eqnarray}
\label{1d}
-\beta f_0^{(l)} \left(\beta J_0^{(l)},\beta h^{(l)}\right)=\log\left\{e^{\beta J_0^{(l)}}\cosh(\beta h^{(l)})+
\left[e^{2\beta J_0^{(l)}}\sinh^2(\beta h^{(l)})+e^{-2\beta J_0^{(l)}}\right]^{\frac{1}{2}}\right\},
\end{eqnarray} 
\begin{eqnarray}
\label{1d1}
m_0^{(l)} (\beta J_0^{(l)},\beta h^{(l)})=\frac{e^{\beta J_0^{(l)}}\sinh(\beta h^{(l)})}
{\left[e^{2\beta J_0^{(l)}}\sinh^2(\beta h^{(l)})+e^{-2\beta J_0^{(l)}}\right]^{\frac{1}{2}}},
\end{eqnarray} 
\begin{eqnarray}
\label{1d2}
\tilde{\chi}_0^{(l)} (\beta J_0^{(l)},\beta h^{(l)})=\frac{e^{-\beta J_0^{(l)}}\cosh(\beta h^{(l)})}
{\left[e^{2\beta J_0^{(l)}}\sinh^2(\beta h^{(l)})+e^{-2\beta J_0^{(l)}}\right]^{\frac{3}{2}}}.
\end{eqnarray} 
Therefore, the self-consistent system (\ref{THEO})-(\ref{THEO1}) for the $n$ communities becomes
\begin{eqnarray}
\label{1d3}
m^{(\Sigma;l)}=\frac{e^{\beta J_0^{(\Sigma;l)}}\sinh(\beta H^{(\Sigma;l)}+\beta h^{(l)})}
{\left[e^{2\beta J_0^{(\Sigma;l)}}\sinh^2(\beta H^{(\Sigma;l)}+\beta h^{(l)})+e^{-2\beta J_0^{(\Sigma;l)}}\right]^{\frac{1}{2}}},
\end{eqnarray} 
\begin{eqnarray}
\label{1d4}
H^{(\Sigma;l)}=\sum_k\beta J^{(\Sigma;l,k)}m^{(\Sigma;k)}.
\end{eqnarray} 
For the symmetric case, from Eqs. (\ref{AF4}) and (\ref{AF5})
we obtain that the $n$ communities undergo a second order transition
at a critical temperature given either by
\begin{eqnarray}
\label{1d5}
\left(\beta_c^{(\Sigma)} J^{(\Sigma;1,2)}-\beta_c^{(\Sigma)} J^{(\Sigma;1,1)}\right)e^{2\beta_c^{(\Sigma)} J_0^{(\Sigma)}}=-1,
\end{eqnarray}
or by
\begin{eqnarray}
\label{1d6}
\left(\left(n-1\right)\beta_c^{(\Sigma)} J^{(\Sigma;1,2)}+\beta_c^{(\Sigma)} J^{(\Sigma;1,1)}\right)
e^{2\beta_c^{(\Sigma)} J_0^{(\Sigma)}}=1,
\end{eqnarray}
where we have used, from Eq. (\ref{1d2}), $\tilde{\chi}_0(\beta J_0;0)=\exp(2\beta J_0)$.
\end{widetext}

\section{Application to percolation}
\subsection{An effective percolation theory}
The key point of our approach is a mapping
that maps the original random model onto a non random one.
In turn this mapping is based on the so called high temperature expansion
of the free energy which, in a suitable region of the phase
diagram that we call P, converges. The boundary of the P region
is established by the critical condition (\ref{THEOs2}).
Within a physical picture, in Sec. IIIB we were mainly concerned with 
the critical temperature. However, it should be clear that the
criticality condition can be expressed in terms of any of the
parameters entering Eq. (\ref{THEOs2}). In particular, for $\Sigma=$F
and non negative couplings $\beta J^{(\mathrm{F};l,k)}\geq 0$ and 
$J_0^{(l,k)}\geq 0$, 
it is of remarkable interest to study the criticality condition in the
limit of zero temperature as well as the behavior of the magnetizations
$m^{(\mathrm{F};l)}$ near this boundary.
Recalling the definition of the set of bonds $\Gamma_0^{(l)}$ and $\Gamma_0^{(l,k)}$ 
(we find it convenient
to define also $\Gamma_0^{(l,l)}\defi\Gamma_0^{(l)}$ and $\Gamma_0\defi\cup_{l',k'}\Gamma_0^{(l',k')}$),
and exploiting $\beta J^{(\mathrm{F};l,k)}\to c^{(l,k)}$, for $\beta\to\infty$,
it is easy to see that in the limit $\beta\to\infty$ the criticality condition (\ref{THEOs2}) amounts to
\begin{eqnarray}
\label{P1}
\det \left(\bm{1}-\bm{\mathcal{E}_0}\cdot\bm{c}\right)=0, \quad \left\{c^{(l,k)}\geq 0\right\} 
\end{eqnarray}
%\begin{eqnarray}
%\label{P1}
%\det \left(\bm{B}\left(\left\{c^{(l,k)}\right\}\right)\right)=0, \quad \left\{c^{(l,k)}\geq 0\right\} 
%\end{eqnarray}
%where the $n\times n$ matix $\bm{B}$ is given by
%\begin{eqnarray}
%\label{P2}
%\bm{B}=\bm{1}-\bm{\mathcal{E}_0}\cdot\bm{c},
%\end{eqnarray}
where $\bm{c}$ is the matrix of the added connectivities $c^{(l,k)}$,
and the $n\times n$ matrix $\bm{\mathcal{E}_0}=\bm{\mathcal{E}_0}(\Gamma_0)$ is given by
%\begin{eqnarray}
%\label{P2}
%&& B^{(l,k)}\defi
%\delta_{l,k} - e_\infty^{(l,l)}\left(\Gamma_0^{(l)}\right)c^{(l,k)},
%\end{eqnarray}
%if $J_0^{(l,k)}=0$ for any $l\neq k$, whereas in the general case
%\begin{eqnarray}
%\label{P3}
%B^{(l,k)}&\defi &
%\delta_{l,k} - \sum_m  
%e_\infty^{(l,m)}\left(\Gamma_0\right)c^{(m,k)},
%\end{eqnarray}
\begin{eqnarray}
\label{P2d}
\mathcal{E}_0^{(l,l)}=\lim_{\beta\to+\infty} \tilde{\chi}_0^{(l,l)}\left(\beta J_0^{(l)};0\right),
\end{eqnarray}
if $J_0^{(l,k)}=0$ for any $l\neq k$, whereas in the general case
\begin{eqnarray}
\label{P3d}
\mathcal{E}_0^{(l,k)}=
\lim_{\beta\to+\infty} \tilde{\chi}_0^{(l,k)}\left(\left\{\beta J_0^{(l',k')}\right\};0\right).
\end{eqnarray}

In the above equations it is understood that we are considering only graphs
$\Gamma_0$ such that the pure model has no finite critical temperature, \textit{i.e.}, it is only $\beta_{c0}=\infty$.
In fact, if this is not the case, as occurs for instance if $\Gamma_0$ is the $d_0$ dimensional lattice with
$d_0\geq 2$, the P region is shrunk to the single trivial point $c^{(l,k)}\equiv 0$ and we can
take effectively $\mathcal{E}_0^{(l,k)}=+\infty$ (this issue will become clearer soon, see also Appendix B).
More precisely, if $\beta_{c0}<\infty$ or even if $\beta_{c0}=0$, as occurs in scale free networks
with a power law exponent $\gamma\leq 3$ \cite{Goltsev}, one should use Eqs. (\ref{P2d}) and (\ref{P3d}) keeping $N$ finite. 

A solution $\{c_c^{(l,k)}\}$ of Eq. (\ref{P1}) represents the exact critical values (or percolation threshold) 
of the set of the connectivities over which
a giant connected component exists. 
Note that Eq. (\ref{P1}) in general is a non linear equation in the $n^2$ unknown connectivities
$\{c{(l,k)}\}$. Therefore, in general, for $n\geq 2$, there are infinite solutions as
$\infty^{n^2}$. This degeneracy of Eq. (\ref{P1}) reflects the fact that given $n$ communities,
we can realize a connected cluster by placing the bonds in 
many ways, as \textit{e.g.}, either over a single community, or between 2 or more communities. 
However, this high degeneracy can be partially removed, and Eq. (\ref{P1}) simplified,
when there are special symmetries among the connectivities.
In particular, the symmetric case ($J_0^{(l,k)}=0$ for $l\neq k$, $c^{(l,l)}=c^{(1,1)}$ for any $l$ and
$c^{(l,k)}=c^{(1,2)}$ for any $l\neq k$) constitutes a simplified situation
in which the determinant in Eq. (\ref{P1}) can be explicitly calculated as in (\ref{AF3}).
In the symmetric case, with the further simplification $J_0^{(l,l)}=J_0$,
Eq. (\ref{P1}) gives the following solution splitted into two branches
\begin{eqnarray}
\label{P4}
\left(c_c^{(1,2)}-c_c^{(1,1)}\right)
\lim_{\beta\to+\infty}\tilde{\chi}_0\left(\beta J_0;0\right)=-1, %\quad c_c^{(1,2)}\leq c_c^{(1,1)}
\end{eqnarray}
and
\begin{eqnarray}
\label{P5}
\left(\left(n-1\right)c_c^{(1,2)}+c_c^{(1,1)}\right)
\lim_{\beta\to+\infty} \tilde{\chi}_0\left(\beta J_0;0\right)=1. %\quad c_c^{(1,2)}+c_c^{(1,1)}\geq 0,
\end{eqnarray}
Notice that the solution with $c_c^{(1,2)}=c_c^{(1,1)}=0$ may occur only for a divergent susceptibility.
In particular, when there are no short-range connections, \textit{i.e.} $J_0=0$, 
we have $\tilde{\chi}_0\left(\beta J_0=0;0\right)=1$ so that
Eqs. (\ref{P4}) and (\ref{P5}) give the surfaces
\begin{eqnarray}
\label{P6}
c_c^{(1,2)}=c_c^{(1,1)}-1, \quad c_c^{(1,1)}\geq 1
\end{eqnarray}
and, for $n\geq 2$
\begin{eqnarray}
\label{P7}
c_c^{(1,2)}=\frac{1-c_c^{(1,1)}}{n-1}, \quad c_c^{(1,1)}<1,
\end{eqnarray}
where we have taken into account that the connectivities are defined non negative.

Similarly, one can study the relative size $\{s^{(l)}\}$ of the giant connected component by 
studying, for $\Sigma=F$ and for non negative short-range couplings, 
the self-consistent system (\ref{THEO})-(\ref{THEO1}) in the limit $\beta \to+\infty$,
where, again, we exploit $\beta J^{(\mathrm{F};l,k)}\to c^{(l,k)}$, for $\beta\to\infty$.
In the case of $J_0^{(l,k)}=0$ for $l\neq k$, we get
\begin{eqnarray}
\label{P8}
s^{(l)}=s_0^{(l)}\left(\sum_k c^{(l,k)}s^{(k)}\right),
\end{eqnarray}
where 
\begin{eqnarray}
\label{P8b}
s_0^{(l)}\left(\sum_k c^{(l,k)}s^{(k)}\right)\defi m_0^{(l)}\left(\lim_{\beta\to\infty}\beta J_0^{(l)};
\sum_k c^{(l,k)}s^{(k)}\right).
\end{eqnarray}
%Here $s^{(l)}$ represents the relative size of the giant connected component (if any)
%of the $l$-th community in the presence of the connectivities $\{c^{(l,k)}\}$, and $s_0^{(l)}\left(c\right)$
%represents the relative size of the giant connected component (if any) of the underlying
%graph $(\mathcal{L}_0^{(l)},\Gamma_0^{(l)})$ in the presence of 
%a number of uniformly distributed random bonds added only over $(\mathcal{L}_0^{(l)},\Gamma_0^{(l)})$ 
%and having a generic average connectivity $c$
%(not to be confused with $c^{(l,l)}$).
%In the general case where $J_0^{(l,k)}\neq 0$ for some $l\neq k$, we have to consider
%the zero temperature limit of the self-consistent system (\ref{THEOG}) and we get
whereas in the general case, from Eq. (\ref{THEOG}) we get
\begin{eqnarray}
\label{P9}
s^{(l)}=s_0^{(l)}\left(\left\{\sum_k c^{(l',k)}s^{(k)}\right\}\right),
\end{eqnarray}
where now 
\begin{eqnarray}
\label{P9b}
&&s_0^{(l)}\left(\left\{\sum_k c^{(l',k)}s^{(k)}\right\}\right)\defi\nonumber \\
&& m_0^{(l)}\left(\left\{\lim_{\beta\to\infty}\beta J_0^{(l',k')}\right\};\left\{\sum_k c^{(l',k)}s^{(k)}\right\}\right).
\end{eqnarray}
%$s_0^{(l)}(\{c_{l'}\})$
%represents the relative size of the giant connected component (if any) of the underlying
%graph $(\mathcal{L}_0^{(l)},\Gamma_0^{(l)})$ in the presence of: an added number
%of random bonds with average connectivity $c_l$, and in the
%presence of the other communities $(\mathcal{L}_0^{(l')},\Gamma_0^{(l')})$
%which in turn are each in the presence of added random bonds with average connectivity $c_l'$.
Note that in the above equations $c^{(l,k)}s^{(k)}$ represents a fraction of the connectivity $c^{(l,k)}$,
being $0\leq s^{(k)}\leq 1$.
Note also that, by virtue of some non zero short-range couplings $J_0^{(l,k)}$ with $l\neq k$,
and $J_0^{(l,h)}$ with $l\neq h$ and $h\neq k$,
the relative sizes of the giant connected component of the two communities $l$ and $h$, $s_0^{(l)}$ and 
$s_0^{(h)}$ (the pure model),
influence each other also if there is no short-cut among the two communities, $J_0^{(l,h)}=0$ .
 
When we are far from the symmetric case, the self-consistent systems (\ref{P8}) or (\ref{P9}) in general may admit
many stable solutions, \textit{i.e.}, solutions $\{s^{(l)}\}$ such that the lhs of Eq. (\ref{P1}) is positive.
The relative probabilities among these solutions can be calculated as $\exp(-L)$ 
from the expressions (\ref{THEO10}) or (\ref{THEO12G}) for the Landau free energy density
in the limit $\beta\to+\infty$. 
In the case of $J_0^{(l,k)}=0$ for $l\neq k$, up to a constant term, we get
\begin{widetext}
\begin{eqnarray}
\label{P10}
L\left(s^{(1)},\ldots,s^{(n)}\right)=\sum_l\alpha^{(l)}
L_0^{(l)}\left(\sum_k c^{(l,k)}s^{(k)}\right)
%\nonumber \\ &+& 
+\sum_{l,k} \frac{\alpha^{(l)}c^{(l,k)}s^{(l)}s^{(k)}}{2},
\end{eqnarray} 
where
\begin{eqnarray}
\label{P11}
L_0^{(l)}\left(\sum_k c^{(l,k)}s^{(k)}\right)
\defi \lim_{\beta\to+\infty}\left[\beta f_0^{(l)}\left(\beta J_0^{(l)};\sum_k c^{(l,k)}s^{(k)}\right)
%\nonumber \\ &-&
-\lim_{N\to\infty}
\frac{1}{N\alpha^{(l)}}\sum_{(i,j)\in\Gamma_0^{(l)}}\log\left[\cosh\left(\beta J_{0;(i,j)}^{(l)}\right)\right]\right].
\end{eqnarray} 
%The term $-L_0^{(l)}$, as a function of a genric $c$, 
%is nothing else than the logarithm of a degenerate partition function 
%of the $l$-th community as follows
%\begin{eqnarray}
%\label{P12}
%L_0^{(l)}\left(c\right)=-\lim_{N\to\infty}
%\frac{1}{N\alpha^{(l)}}\log\left[\sum_{\{\sigma_i\},i\in\mathcal{L}_0^{(l)}}e^{c\sum_{i\in\mathcal{L}_0^{(l)}}\sigma_i }
%\prod_{(i,j)\in\Gamma_0^{(l)}}\left(1+\sigma_i\sigma_j\right)\right].
%\end{eqnarray} 
Notice that in defining $L_0^{(l)}$ in Eq. (\ref{P11}) we have subtracted the trivial and
divergent terms $\log\left[\cosh\left(\beta J_{0;(i,j)}^{(l)}\right)\right]$
which, being constants, do not modify the relative probabilities.
Similarly, in the general case we have
\begin{eqnarray}
\label{P13}
L\left(s^{(1)},\ldots,s^{(n)}\right)&=&\sum_{l,k} \frac{\alpha^{(l)}c^{(l,k)}s^{(l)}s^{(k)}}{2}+
L_0\left(\left\{\sum_k c^{(l',k)}s^{(k)}\right\}\right)
\end{eqnarray} 
where, 
\begin{eqnarray}
\label{P14}
L_0\left(\left\{\sum_k c^{(l,k)}s^{(k)}\right\}\right)
&\defi& \lim_{\beta\to+\infty}\left[\beta f_0\left(\left\{\beta J_0^{(l')}\right\};
\left\{\sum_k c^{(l',k)}s^{(k)}\right\}\right)
\right. \nonumber \\ &-& \left.
\lim_{N\to\infty}\frac{1}{N}\sum_{(i,j)\in\cup_{l,k}\Gamma_0^{(l,k)}}
\log\left[\cosh\left(\beta J_{0;(i,j)}^{(l,k)}\right)\right]\right],
\end{eqnarray} 
%\begin{eqnarray}
%\label{P15}
%L_0\left(c_1,\ldots,c_n\right)=-\lim_{N\to\infty}
%\frac{1}{N}\log\left[\sum_{\{\sigma_i\},i\in\cup_l\mathcal{L}_0^{(l)}}e^{\sum_l c_l\sum_{i\in\mathcal{L}_0^{(l)}}\sigma_i }
%\prod_{(i,j)\in\cup_{l,k}\Gamma_0^{(l,k)}}\left(1+\sigma_i\sigma_j\right)\right].
%\end{eqnarray} 
\end{widetext}

Of course, among all the stable solutions of the self-consistent system (\ref{P8}) or (\ref{P9})
the leading solution will be given by looking at the absolute minimum of $L$.
It is interesting to observe that the term $-L_0$, as a function of a generic $c$, 
is nothing else than the logarithm of a degenerate partition function. We report its expression in Appendix A. 

The self-consistent systems (\ref{P8}) or (\ref{P9}) are exact in the P region,
that is for all values of the vector $\{c^{(l,k)}\}$ below the leading (\textit{i.e.}, minimal) critical surface
$\{c_c^{(l,k)}\}$, however, again, the limits $\{c^{(l,k)}\to 0^+\}$
and $\{c^{(l,k)}\to\infty\}$ will be exact independently of the critical
surface $\{c_c^{(l,k)}\}$. Under these limitations, our result for percolation
constitutes a generalization to generic $n$ 
of the approach developed in \cite{Newman0,Newman} which was built for $n=1$.

We end this paragraph by giving the meaning of the above expressions directly
in terms of graph theory. Concerning $s_0^{(l)}$ and $s^{(l)}$ they clearly represent 
the fraction of the sites belonging to the giant connected component (if any)
in the $l$-th community. More precisely, in the most general case we have
\begin{eqnarray}
\label{Perco}
s_0^{(l)}\left(\left\{0\right\}\right)=\lim_{N\to\infty}\frac{\sum_{i\in\mathcal{L}_0^{(l)}} n_i^{(l)}}{N\alpha^{(l)}}, 
\end{eqnarray}
where $n_i^{(l)}=0,1$ if respectively the vertex $i\in\mathcal{L}_0^{(l)}$ 
belongs or not to a connected cluster with the property to be giant (if any) and having bonds belonging  
to the given initial set of bonds $\cup_{l',k'}\Gamma_0^{(l',k')}$.
Similarly, $s_0^{(l)}\left(\left\{c^{(l',k')}\right\}\right)$ and $s^{(l)}$
have analogous expressions as Eq. (\ref{Perco}) and
represent the fraction of sites of the $l$-th community 
belonging to a connected component with the property to be giant (if any) 
and having bonds belonging to the union of the initial 
set of bonds $\cup_{l',k'}\Gamma_0^{(l',k')}$ and the other bonds randomly spread according 
to the set of the given added connectivities $\left\{c^{(l',k')}\right\}$.
Note that, by definition, the only difference between $s_0^{(l)}\left(\left\{c^{(l',k')}\right\}\right)$ and $s^{(l)}$
is that 
in the latter case we have to average over the different realizations of the graph
with the distribution (\ref{PP}), while for the former, 
as for the case of thermodynamics, only the average values of the connectivities are taken into account.
Concerning instead the meaning of the matrix $\mathcal{E}_0^{(l,k)}$ we have (see Appendix B for details)
\begin{eqnarray}
\label{Perco3}
\mathcal{E}_0^{(l,l)}=\lim_{N\to\infty}
\sum_{i\in\mathcal{L}_0^{(l)}}\frac{\mathcal{N}_i^{(l)}}{N\alpha^{(l)}}, 
\end{eqnarray}
where $\mathcal{N}_i^{(l)}$ is defined as the number of vertices belonging to the $l$-th community (including $i$ itself) 
which are reachable from the site $i\in\mathcal{L}_0^{(l)}$ by at least one path of connected vertices.   
We recover hence immediately that, \textit{e.g.}, for the case of no short-range coupling $J_0\equiv 0$,
we have $\mathcal{N}_i^{(l)}\equiv 1$ and then $\lim_{\beta\to\infty}\tilde{\chi}_0^{(l,l)}\left(\beta J_0^{(l)};0\right)=1$.
Similarly for the case in which $(\mathcal{L}_0^{(l)},\Gamma_0^{(l)})$ is the one dimensional chain we have
$\mathcal{N}_i\equiv N$ and then $\lim_{\beta\to\infty}\tilde{\chi}_0^{(l,l)}\left(\beta J_0^{(l)};0\right)=+\infty$.
Finally, for $l\neq k$, by defining 
$\mathcal{N}_i^{(l,k)}$, with $i$ belonging to the $l$-th community, 
as the number of vertices belonging to the $k$-th community
reachable by at least one chain of bonds from the vertex $i$,  
we have the following expression analogous to Eq. (\ref{Perco3})
\begin{eqnarray}
\label{Perco4}
\mathcal{E}_0^{(l,k)}
=\lim_{N\to\infty}\sum_{i\in\mathcal{L}_0^{(l)}}\frac{\mathcal{N}_i^{(l,k)}}{N\alpha^{(l)}}.
\end{eqnarray}
Note that now it can be also $\mathcal{N}_i^{(l,k)}=0$, while, by definition $\mathcal{N}_i^{(l,l)}\geq 1$. 
Furthermore it holds the following balance
\begin{eqnarray}
\label{Perco5}
\alpha^{(l)}\mathcal{E}_0^{(l,k)}=
\alpha^{(k)}\mathcal{E}_0^{(k,l)}.
\end{eqnarray}

By using the obvious generalization of Eqs. (\ref{Perco3}) and (\ref{Perco4}),
we can define a quantity similar to $\mathcal{E}_0^{(l,k)}$, $\mathcal{E}^{(l,k)}$, to
include the presence of added bonds randomly spread according to the measure (\ref{PP1})
with average connectivities $\{c^{(l,k)}\}$, and 
from Eq. (\ref{THEOsusc}) in the zero temperature limit we get (in matrix form): 
\begin{eqnarray}
\label{THEOsusc01}
\bm{\mathcal{E}}=\left(\bm{1}-\bm{\mathcal{E}_0}\cdot\bm{c}\right)^{-1}
\cdot \bm{\mathcal{E}_0}.
\end{eqnarray}
Eq. (\ref{THEOsusc01}) tells us how $\bm{\mathcal{E}}$ changes as we vary the $\{c^{(l,k)}\}$,
being an exact equation as the $\{c^{(l,k)}\}$ belong to the P region.

\subsection{Percolation threshold \textit{vs} fractal dimension}
Let us come back, for the moment being, to the case $n=1$ and 
let us consider the simplest model: the VB model (see Sec. IV). In this case
Eq. (\ref{P1}) gives immediately the critical value $c_c=1$,
which is the very long known value of the percolation threshold
for the Poissonian graph \cite{Classical}. On the other hand,
if we have a $d_0$-dimensional underlying pure graph $(\mathcal{L}_0,\Gamma_0)$ 
and $d_0\geq 1$, whatever $(\mathcal{L}_0,\Gamma_0)$ may be complicated,
we always have $\mathcal{E}_0=\mathop{O}(N)$ or, equivalently,
in the limit $\beta \to+\infty$ for the susceptibility of the 
pure system we have $\tilde{\chi}_0\to+\infty$, so that
for the critical value we get always the obvious result $c_c=0$.
To have a critical value $c_c$ strictly in the range $(0,1)$ 
it is necessary that the underlying graph $(\mathcal{L}_0,\Gamma_0)$
has a dimension lower than 1: $d_0<1$. For example one can consider the case
in which the graph $(\mathcal{L}_0,\Gamma_0)$ is made up of finite units
as dimers, triangles, etc.. For instance, in the Ref. \cite{SW} we have studied
the case of dimers for which we have $\tilde{\chi}_0(\beta J_0;0)=(\exp(\beta J_0))/\cosh(\beta J_0)$.
Therefore in the limit $\beta \to +\infty$ we get $\tilde{\chi}_0\to 2$
or, equivalently, $\mathcal{E}_0=2$, 
which inserted in Eq. (\ref{P1}) gives the result $c_c=1/2$.
This result is still trivial because in this example we have $d_0=0$.  
However, there is no limitation in the choice of the graph
$(\mathcal{L}_0,\Gamma_0)$; all the theory holds for any arbitrary 
graph $(\mathcal{L}_0,\Gamma_0)$ which in particular can be some quenched
graph obtained by removing randomly a sufficient number of bonds 
from another initial regular graph so that the final graph $(\mathcal{L}_0,\Gamma_0)$ will have
a non trivial value of $\mathcal{E}_0$ corresponding to a fractal dimension $d_0\in (0,1)$ \cite{Mandelbrot}.
Similarly, as already stressed in Sec. II for generic $n$, for any $l$ and any $k$, the symbol
$J_0^{(l,k)}$ is actually a short notation to indicate all the short-range
couplings connecting the $l$-th and the $k$-th community:
$J_0^{(l,k)}=\{J_{0;(i,j)}^{(l,k)}\}$, $i\in\mathcal{L}_0^{(l)}$, $i\in\mathcal{L}_0^{(k)}$;
so that the graphs $\{(\mathcal{L}_0^{(l,k)},\Gamma_0^{(l,k)})\}$ are completely arbitrary
with a non trivial percolation threshold surface coming from Eq. (\ref{P1}).
Typically, scale free networks, as the Internet, own fractal properties \cite{FractalInternet},
however in this case $d_0\to\infty$  in the thermodynamic limit, 
so that the percolation threshold in these networks becomes zero. 
%as it will become clear in the next paragraph, their own characteristic time to exchange a unit of
%information is zero. 

\subsection{When two given communities do communicate}
The answer to the fundamental question ``when and how much two given communities communicate'' is encoded
in $\tilde{\chi}^{(l,k)}_0$ and $\tilde{\chi}^{(l,k)}$ which at $T=0$ means
$\mathcal{E}^{(l,k)}_0$ and $\mathcal{E}^{(l,k)}$, respectively.
Given the arbitrary pure graph $(\mathcal{L}_0,\Gamma_0)$, 
and some community structure assignment which splits the set of bonds $\Gamma_0$
in $n(n-1)/2$ sets, $\Gamma_0=\cup_{l,k}\Gamma_0^{(l,k)}$, at $T=0$, 
in the pure graph the communities $l$ and $k$ communicate
if and only if $\mathcal{E}_0^{(l,k)}\neq 0$.
We can understand the communication process as follows. 
%Let us suppose that, at a given initial time, an external magnetic field $h^{(k)}$ is acting 
%only on the $k$-th community.    
If suddenly, at a given initial time, appears an external magnetic field $h^{(k)}$,
which acts uniformly only on the spins of the $k$-th community, 
these spins are forced to change and, as a consequence,
all the spins of the other communities will have to suitably change in order to reach the new
equilibrium state. At finite temperature 
the new equilibrium state will be reached after a relaxation time $\tau$ which grows with the size of the system.
However in the limit $T\to 0$ there is no thermal dissipation and 
the spin changes take place instantaneously. Therefore, if the external field $\beta h^{(k)}$
changes with time with a constant velocity, say $v^{(k)}$, from the relation $\delta m^{(l)}\simeq\tilde{\chi}^{(l,k)}\delta(\beta h^{(k)})$ 
extrapolated at $T=0$, we see that 
the characteristic time $t_0^{(l,k)}$ to transmit a unit of information from the community $k$ to the
community $l$ in the pure model will grow as 
\begin{eqnarray}
\label{Percotime0}
v^{(k)}t_0^{(l,k)}=\left(\mathcal{E}_0^{(l,k)}\right)^{-1}.
\end{eqnarray}
Similarly, in the random model having a matrix of added connectivities $\bm{c}$, the communities
$l$ and $k$
communicate with a characteristic time given by
\begin{eqnarray}
\label{Percotime1}
v^{(k)}t^{(l,k)}=\left(\mathcal{E}^{(l,k)}\right)^{-1}.
\end{eqnarray}
From Eq. (\ref{THEOsusc01}) we see that 
in the pure model, if $\mathcal{E}^{(l,k)}_0=0$, the two communities $l$ and $k$ cannot communicate ($t^{(l,k)}_0\to\infty$),
but for any arbitrary small $c^{(l,k)}>0$ they communicate and the characteristic time decays with $\bm{c}$
approximately as (recall that for any $l$ is always $\mathcal{E}^{(l,l)}_0\geq 1$)
\begin{eqnarray}
\label{Percotime2}
v^{(k)}t^{(l,k)}\simeq\left(\mathcal{E}^{(l,k)}_0+\mathcal{E}^{(l,l)}_0\mathcal{E}^{(k,k)}_0c^{(l,k)}\right)^{-1}.
\end{eqnarray}
Similar relations hold also at finite $T$ provided the velocity of the signal, $v^{(k)}$, be sufficiently small
so that $\tau\ll t_0^{(l,k)}$ or $\tau\ll t^{(l,k)}$.

In general if, due to the birth of some giant connected component in the pure model, one has 
$\mathcal{E}^{(l,k)}_0\to\infty$ in the thermodynamic limit, correspondingly we have $t_0^{(l,k)}\to 0$, and then also
$t^{(l,k)}\to 0$; \textit{i.e.}, the communities communicate instantaneously (they percolate). 
However, in the random model, even if $\mathcal{E}^{(l,k)}_0$ is finite, provided non zero, 
when $\bm{c}$ approaches the percolation threshold surface $\bm{c}_c$, 
given by Eq. (\ref{P1}), then we have $t^{(l,k)}\to 0$. 
Of course intermediate situations will give rise
to finite values of $t^{(l,k)}$. In Eq. (\ref{THEOsusc01}) the matrix $\bm{\mathcal{E}_0}$ represents
an input data. In general, it can be sampled efficiently by simple simulated annealing procedures by using Eq. (\ref{P3d}),
since the problem is mapped to an unfrustrated Ising model ($\beta J_0^{(l,k)}\geq 0$). 

We recall that there is no limitation in the choice of the graph $(\mathcal{L}_0,\Gamma_0)$
and that the statistical mechanical framework provided through 
Eqs. (\ref{P2d},\ref{P3d}) and Eqs. (\ref{Perco})-(\ref{Perco5})
is complete and exact. Note in particular that two given communities $l$ and $k$,
when immersed in a context of more communities, can communicate even
if there is no bond between them, $\Gamma_0^{(l,k)}=\{0\}$, due to the presence of one (or more) chain
of communities other than $l$ and $k$ that start from $l$ and arrive to $k$ through a sequence of, say $m$, sets
$\Gamma_0^{(l,l_1)},\ldots,\Gamma_0^{(l_m,k)}$. In other words, if we activate all the couplings of the set $\Gamma_0$,
$\{\beta J_0^{(l',k')}>0\}$, Eq.(\ref{P3d}) takes into account that 
a message can go through any of the different paths of communities. 
We point out also that, in the pure model, having some bonds between the $l$-th and $k$-th communities 
does not ensure that the condition $\mathcal{E}^{(l,k)}_0>0$ be satisfied.
From Eq. (\ref{Perco4}) we see in fact that for having $\mathcal{E}^{(l,k)}_0>0$ it is necessary that
the number of paths between the $l$-th and the $k$-th communities be at least of order $N$. 
Note also that such a requirement does not exclude the possibility 
that even a single bond between the two communities be enough, provided this bond has
a very high betweenness.
It should be then clear that an analysis of the given graph $(\mathcal{L}_0,\Gamma_0)$ based
only on simple algorithms making use of the adjacency matrix, and that are therefore local,
can never capture global features as the ones we have elucidated above and that determine the
real communication properties of the network.

\subsection{Percolation \textit{vs} community structure}
%We end this section observing that, quite interestingly - as an inverse problem - by varying
%the whole set of the non zero short-range bonds through which a giant connected
%component may form, one can investigate the structure of the communities.
The analysis of the previous paragraph suggests also a possible criterion to detect community structures.
%The understanding of the community structure from real data is in fact 
%a major problem which in recent years has attracted much attention. Many methods have been proposed
%and special progresses have been made by mapping the problem for identifying community structures to
%optimization problems \cite{Girvan,Caldarelli,Jorg} or by looking for $k-$clique subgraphs \cite{Palla}.
Our idea to detect community structures comes from the physical picture of percolation. 
%and its phylosophy is maybe more related to \cite{Palla}.
Given an arbitrary graph $(\mathcal{L}_0,\Gamma_0)$, and an hypothetical number of communities $n$,
we look for the partition $(\mathcal{L}_0=\cup_{l=1}^n\mathcal{L}_0^{(l)},\Gamma_0=\cup_{l,k}^{n}\Gamma_0^{(l,k)})$ 
such that the resulting community's structure has minimal communication, \textit{i.e.}, 
according to the previous paragraph, we look for the minimization of the sum of the non diagonal
matrix elements of the matrix $\bm{\mathcal{E}_0}$. It is possible to make more precise
this criterion by considering the modularity introduced by Newman and Girvan.
In \cite{Girvan}, given an arbitrary graph $(\mathcal{L}_0,\Gamma_0)$, 
one introduces a ``measure'' $Q_1=Q_1(\Gamma_0)$ - known as modularity - 
of the quality of assignment of vertices into communities. It is defined as
\begin{eqnarray}
\label{Perco7}
Q_1=\sum_{l}\left[e_1^{(l,l)}-\left(a_1^{(l)}\right)^2\right],
\end{eqnarray}
where: $e_1^{(l,k)}$, with $l\neq k$, is  
the fraction of all bonds connecting the two communities $l$ and $k$, 
$e_1^{(l,l)}$
is the fraction of bonds falling inside the community $l$, and
$a_1^{(l)}$ is defined as $a_1^{(l)}\defi\sum_{k}e_1^{(l,k)}$, the fraction of all bonds
having one or two ends in the community $l$.
The term $(a_1^{(l)})^2$ in Eq. (\ref{Perco7})
represents the expected fraction of bonds falling inside the community $l$ when
their ends are connected randomly. 
Thanks to the presence of the term $(a_1^{(l)})^2$ in Eq. (\ref{Perco7}), 
$Q_1$ gives measure 0 when one considers
the trivial case in which $\Gamma_0$ is a single community ($n=1$), and partitions
that maximize $Q_1$ correspond to best community structures.
Nothing however avoids us to define another similar measure which takes into account not only bonds,
but also, for example, paths of two consecutive bonds. In general we can define 
\begin{eqnarray}
\label{Perco8}
Q_h=\sum_{l}\left[e_h^{(l,l)}-\left(a_h^{(l)}\right)^2\right],
\end{eqnarray}
where now $e_h^{(l,k)}$ for $l\neq k$ is the fraction of all paths of length not greater than $h$
connecting the two communities $l$ and $k$, 
$e_h^{(l,l)}$ is the fraction of all paths of length not greater than $h$ whose both ends are in the $l$-th community,
and $a_h^{(l)}\defi\sum_{k}e_h^{(l,k)}$ is the total number of paths of length not greater than $h$ 
having one or both the ends in the community $l$. 
Again we have that its square represents the expected fraction of paths of length not greater than
$h$ having both ends inside the community $l$ when they are connected randomly, 
and makes the measures (\ref{Perco8}) non trivial.

When we optimize $Q_1$ with respect to all the possible ways of assignment of vertices into communities,
we are looking for the best case (or the cases) in which there are ``few'' bonds between different communities
and ``many'' inside the same community.  
Similarly, when we optimize $Q_h$ with respect to all the possible ways of assignment of vertices into communities,
we are looking for the best case (or the cases) in which there are ``few'' paths of length $\leq h$ whose ends arrive in
different communities and ``many'' paths whose ends arrive in the same community. 
Clearly, the bigger is $h$, the stricter 
and demanding is our definition of $Q_h$. 
Consider for example that we are looking for an assignment in
two communities, $n=2$. An assignment optimal with respect to $Q_1$ will give typically 
many bonds inside each communities and only a few between the two. On the other hand, inside for instance 
the community 1, there may be many paths of length not greater than $h-1$ converging at a same vertex inside
the community 1, vertex which in turn is connected by a further bond to the community 2.
Therefore, in general, the above assignment of 2 communities will not be the best assignment with respect to $Q_h$,
and in this example $Q_h$ will be better optimized for just one community, $n=1$. 
In general $Q_h$ will selected assignments such that between any two communities there are either
``few'' bonds or, if there are ``many'' bonds, such bonds must have a small betweenness, and   
the greater is $h$, the smaller will be such allowed betweenness.
More precisely, we see that there are basically two regimes: when $h\ll N/n$, $N/n$ being the typical
size of one community, the probability that a path of length not greater than $h$
leaves the community from which it starts, say the $k$-th one, 
depends only on the local topology near the $k$-th community and on its boundaries with 
the other nearest communities, whereas when $h\gg N/n$ this probability will be highly
affected by the value of $N/n$ and will be very large for small values of $N/n$; \textit{i.e.} for the
case of many communities.
Eventually, in the thermodynamic limit, for $h\to\infty$, $Q_h$ will select a community's assignment such that
for any two communities there is zero betweenness and, in this limit,
the above definition of $e_h^{(l,k)}$, up to a constant factor which is independent of the community's assignment, 
coincides with Eqs. (\ref{Perco3}) and (\ref{Perco4}).
In this limit, in particular, if some percolation takes place, the community's structure
that maximizes $Q_h$ will coincide with the ensemble of the finite and infinite (percolating) 
connected clusters that form in the graph. 

We find that our approach for community detection has some analogies to the random walks approach developed in
\cite{Lambiotte}. Actually, as stressed in \cite{Lambiotte}, there are many different ways to define
dynamical processes (with continuous or discrete time) 
able to probe the community structure of networks having different features. 
In particular, given the adjacency matrix of the given graph, $A_{i,j}$,
%(in our paper this symbol corresponds to $A_{i,j}=c^{(0)}_{i,j}+c_{i,j}$, where $c^{(0)}_{i,j}$ and $c_{i,j}$ 
%are the adjacency matrices of the pure model and the additional connectivities of the and random model, respectively, 
%however for simplicity we can take here $c_{i,j}=0$), 
we can distinguish two family of dynamical processes built through the normalized or unnormalized Laplacian
operator. Our approach belongs to the second family so that uncontrolled large fluctuations are 
involved when sampling the evolution operator $\exp(-At)$ at large times and the method turns out to be 
in principle inefficient. However, as explained before, in the limit $t\to\infty$ our method 
does not need to sample the operator $\exp(-At)$, but only to evaluate the susceptibility
of the unfrustrated Hamiltonian model $H_0$ at zero temperature via simulated annealing. 
As has been pointed out in \cite{Lambiotte}, the methods developed in \cite{Guimera} and \cite{Jorg}
which map the problem of community detection to that of finding the ground state
of a suitable frustrated Potts model, are equivalent to considering the modularity built
through dynamical processes at small but finite $t$ ($\mathop{O}(1)$ in adimensional units).
It is then clear that our proposed method of community detection, although be defined
through an Hamiltonian model too, is completely different from these other Hamiltonian methods. As mentioned
in the introduction the reason is due to the fact that our method is based
on the correlation functions, and not on the energies, and the correlation functions,
even at equilibrium, have a well defined relationship with dynamics.

%%%%%%%%%%%%%%%%%
\begin{widetext}
\section{Mapping to a non random model}
In the Ref. \cite{SW} we have made use of a general mapping: an Ising model 
with random couplings and defined over a random graph is mapped onto a
non random Ising model turning out (in that specific case of just one community)
an Ising model defined over the fully connected graph $\Gamma_f$,
and having only a constant long-range coupling, $\beta J^{(I)}$, 
and a constant short-range coupling, $\beta J_0^{(I)}$, suitable tuned
by the measures of the graph-disorder and the coupling-disorder of the original model via the identities:
$\beta J^{(I)}=\beta J^{(\Sigma)}$ and $\beta J_0^{(I)}=\beta J_0^{(\Sigma)}$.
More precisely, in the Ref. \cite{SW} we saw that the mapping consisted of two steps.
In a first step we map the original model onto a random Ising model with no more graph-disorder
and built over $\Gamma_f$. Then, in a second step we
map the latter model onto an Ising model with no more coupling-disorder.
Due to the independence of the random matrix elements $c_{i,j}^{(l,k)}$ and
of the random couplings $J_{i,j}^{(l,k)}$, it is not difficult to generalize 
the mapping to our case of $n$ communities.
By defining $\tilde{J}_{i,j}^{(l,k)}$ as 
\begin{eqnarray}
\label{Jtilde0}
\tilde{J}_{(i,j)}^{(l,k)}\defi {J}_{(i,j)}^{(l,k)}g_{(i,j)}^{(l,k)}, \quad g_{(i,j)}^{(l,k)}=0,1, 
\end{eqnarray}
after the first step, the free energy (\ref{logZ}) reads as
\begin{eqnarray}
\label{logZG}
-\beta F\defi 
\int d\tilde{\mathcal{P}}\left(\tilde{\bm{J}}\right)
\log\left(\tilde{Z}_{\Gamma_f;\tilde{\bm{J}}}\right),
\end{eqnarray}
where: $\Gamma_f$ is the ``fully connected graph'' obtained
as union of all the fully connected graphs for 
each community $l$ and for each couple of communities $(l,k)$ 
\begin{eqnarray}
\label{Full}
\Gamma_f\defi\cup_l\Gamma_f^{(l)}\cup_{l<k}\Gamma_f^{(l,k)};
\end{eqnarray}
$\tilde{Z}_{\Gamma_f}$ is the partition function of the random Ising model defined
over $\Gamma_f$ with the random Hamiltonian
\begin{eqnarray}
\label{HM2}
\tilde{H}_{\tilde{\bm{J}}} \left(\left\{\sigma_i\right\}\right)
&\defi& -\sum_{(i,j)\in\Gamma_f} \tilde{J}_{(i,j)} \sigma_{i}\sigma_{j}
\nonumber \\
&-&\sum_lh^{(l)}\sum_{i\in\mathcal{L}_0^{(l)}}\sigma_i;
%+\sum_{i=1}^N h_i \sigma_i,
\end{eqnarray} 
and the $\tilde{\bm{J}}$'s are distributed according 
to the measure $d\tilde{\mathcal{P}}\left(\tilde{\bm{J}}\right)$ given by
\begin{eqnarray}
\label{dP}
d\tilde{\mathcal{P}}\left(\tilde{\bm{J}}\right)&\defi& \prod_l\prod_{~~i<j,~i,j\in\mathcal{L}_0^{(l)}} 
d\tilde{\mu}^{(l,l)}\left( \tilde{J}_{i,j}^{(l,l)} \right),
\nonumber \\
&\times&\prod_{l<k}\prod_{~~i\in\mathcal{L}_0^{(l)},j\in\mathcal{L}_0^{(k)}} 
d\tilde{\mu}^{(l,k)}\left( \tilde{J}_{i,j}^{(l,k)} \right),
\end{eqnarray}
with
%\begin{widetext}
\begin{eqnarray}
d\tilde{\mu}_{(i,j)}^{(l,k)}\left(\tilde{J}_{(i,j)}^{(l,k)}\right) = 
\left\{
\begin{array}{l}
\label{sep2}
d\mu_0^{(l,k)}\left(J_{(i,j)}^{(l,k)}\right), \quad \quad \quad \qquad ~ \quad (i,j)\in\Gamma_0^{(l,k)},\\
\label{sep3}
d\mu^{(l,k)}\left(J_{(i,j)}^{(l,k)}\right)p^{(l,k)}\left(g_{(i,j)}^{(l,k)}\right), 
\quad (i,j)\in\Gamma_f^{(l,k)}\backslash \Gamma_0^{(l,k)},
\end{array}
\right.
\end{eqnarray}
$d\mu^{(l,k)}(\cdot)$ and $p^{(l,k)}(\cdot)$ being
the coupling- and bond- (Eq. (\ref{PP1})) measures of the original model introduced in Sec. II, 
and $d\mu_0^{(l,k)}(J_0)/dJ_0$ the delta distribution around the given short-range coupling $J_0^{(l,k)}$.
For the second step we have to use
the general rule of the mapping to map a random Ising model built over a quenched graph
to a non random Ising model (\textit{the related Ising model}, whose quantities we label with a suffix $I$) 
having suitable couplings \cite{MOI}.
Depending on whether we are looking for the solution with label F or the solution
with label SG, the coupling or mapping substitutions are given by 
\begin{eqnarray}
\label{mapG}
\left\{\tanh\left(\beta \tilde{J}_{(i,j)}^{(l,k)}\right)\right\}\to
\left\{\int d\tilde{\mu}\left(\tilde{J}_{(i,j)}^{(l,k)}\right)\tanh^{l_\Sigma}\left(\beta \tilde{J}_{(i,j)}^{(l,k)}\right)\right\},
\end{eqnarray}
where $l_\Sigma=1$ or $2$ for $\Sigma=$F or SG, respectively.
We recall also that there are no intermediate transformations mixing
the F and the SG solution: there are only two physical transformations, be F or SG,
affecting simultaneously all the couplings.
After the second step, the free energy of the original problem reads as 
\begin{eqnarray}
\label{logZG2}
-\beta F&=& \sum_l\sum_{(i,j)\in\Gamma_0^{(l)}}\log\left[\cosh\left(\beta J_0^{(l)}\right)\right]+
\sum_{l<k}\sum_{(i,j)\in\Gamma_0^{(l,k)}}\log\left[\cosh\left(\beta J_0^{(l,k)}\right)\right]+\phi, 
\nonumber \\ &+&
\sum_l\sum_{(i,j)\in\Gamma_f^{(l)}}\int d\tilde{\mu}_{(i,j)}^{(l,k)}\left(\tilde{J}_{(i,j)}^{(l,k)}\right)
\log\left[\cosh\left(\beta J_0^{(l)}\right)\right]+
\sum_{l<k}\sum_{(i,j)\in\Gamma_f^{(l,k)}}\int d\tilde{\mu}_{(i,j)}^{(l,k)}\left(\tilde{J}_{(i,j)}^{(l,k)}\right)
\log\left[\cosh\left(\beta J_0^{(l,k)}\right)\right],
\end{eqnarray}
where $\phi$ is the non trivial part of the free energy, whose density $\varphi$,
(and similarly any correlation function $C$)
in the thermodynamic limit can be calculated through $\varphi_I$ ($C_I$), the free energy
density (the correlation function) 
of the related Ising model having couplings obeying Eq. (\ref{mapG})~\footnote{See the discussion
provided in the last part of Sec. IIIA.1 for clarifying how to use the mapping for
calculating physical correlation functions.}.
%\end{widetext} 

\section{Derivation of the self-consistent equations}
By using the above result, we are now able to derive 
Eqs. (\ref{THEO}-\ref{THEOs2}). To this aim we have to solve the thermodynamics
of the following related Ising model
%\begin{widetext} 
\begin{eqnarray}
\label{HI}
H_I &=& -\sum_l J_{0}^{(I;l,l)}\sum_{(i,j)\in \Gamma_0^{(l)}}\sigma_{i}\sigma_{j}
-\sum_{l<k} J_{0}^{(I;l,k)}\sum_{(i,j)\in \Gamma_0^{(l,k)}}\sigma_{i}\sigma_{j} \\
\nonumber &-&\sum_l J^{(I;l,l)}\sum_{(i,j)\in \Gamma_f^{(l)}}\sigma_{i}\sigma_{j}
-\sum_{l<k} J^{(I;l,k)}\sum_{(i,j)\in \Gamma_f^{(l,k)}}\sigma_{i}\sigma_{j} \\
\nonumber &-& \sum_l h^{(l)}\sum_{i\in\mathcal{L}_0^{(l)}} \sigma_i.  
\end{eqnarray}
Note that the above couplings (uniform within each appropriate set) $J_{0}^{(I;l,k)}$ and $J^{(I;l,k)}$
are arbitrary. In fact, the mapping requires to solve the 
thermodynamics of the related Ising model with arbitrary adimensional couplings
$\beta J_{0}^{(I;l,k)}$ and $\beta J^{(I;l,k)}$ 
and only after to perform the mapping substitutions (\ref{mapG}).
Therefore, as done in the Ref. \cite{SW}, we find it convenient - for physical and conventional reasons -
to consider not the Hamiltonian (\ref{HI}) and the transformations (\ref{mapG}), but the following 
Hamiltonian and transformations
\begin{eqnarray}
\label{HI1}
H_I &=& -\sum_l J_{0}^{(I;l,l)}\sum_{(i,j)\in \Gamma_0^{(l)}}\sigma_{i}\sigma_{j}
-\sum_{l<k} J_{0}^{(I;l,k)}\sum_{(i,j)\in \Gamma_0^{(l,k)}}\sigma_{i}\sigma_{j} \\
\nonumber &-&\sum_l \frac{J^{(I;l,l)}}{N\alpha^{(l)}}\sum_{(i,j)\in \Gamma_f^{(l)}}\sigma_{i}\sigma_{j}
-\sum_{l<k} \frac{J^{(I;l,k)}}{N\alpha^{(l,k)}}\sum_{(i,j)\in \Gamma_f^{(l,k)}}\sigma_{i}\sigma_{j} \\
\nonumber &-& \sum_l h^{(l)}\sum_{i\in\mathcal{L}_0^{(l)}} \sigma_i,
\end{eqnarray}
\begin{eqnarray}
\label{mapG1}
\left\{\tanh\left(\beta \frac{\tilde{J}_{(i,j)}^{(l,k)}}{N_{(i,j)}^{(l,k)}}\right)\right\}\to
\left\{\int d\mu\left(\tilde{J}_{(i,j)}^{(l,k)}\right)\tanh^{l_\Sigma}\left(\beta \tilde{J}_{(i,j)}^{(l,k)}\right)\right\},
\quad N_{(i,j)}^{(l,k)}\defi
\left\{
\begin{array}{l}
1, \qquad ~ ~ \quad (i,j)\in \Gamma_0^{(l,k)} \\
N\alpha^{(l,k)}, \quad (i,j)\in \Gamma_f^{(l,k)}\backslash\Gamma_0^{(l,k)},
\end{array}
\right.
\end{eqnarray}
where we have introduced the coefficient $\alpha^{(l,k)}$, giving
the total number of possible bonds between the $l$-th and the $k$-th community, via
$N^{(l,k)}=\alpha^{(l,k)}N$
%and we have also defined $\alpha^{(l,l)}\defi \alpha^{(l)}$.
\begin{eqnarray}
\label{alphaalpha}
\alpha^{(l,k)} \defi 
\left\{
\begin{array}{l}
\alpha^{(l)}, \quad l=k, \\
\alpha^{(l)}\alpha^{(k)}, \quad l\neq k,
\end{array}
\right.
\end{eqnarray}
where $\alpha^{(l)}$ has been introduced in Sec. II and is related to the size $N^{(l)}$
of the $l$-th community via $N^{(l)}=\alpha^{(l)}N$.

By using now Eq. (\ref{PP1}) in Eqs. (\ref{Jtilde0})-(\ref{sep3}), from Eq. (\ref{mapG1}) 
applied for large $N$, we obtain that, after solving the thermodynamics of the related Ising model with
Hamiltonian (\ref{HI1}), the mapping transformations for any $l,k$ read as
\begin{eqnarray}
\label{mapG2}
%\begin{array}{l}
\beta J^{(I;l,k)} &\to& \frac{\alpha^{(l,k)}}{\alpha^{(k)}}\beta J^{(\Sigma;l,k)}, \\
\label{mapG2b}
\beta J_{0}^{(I;l,k)} &\to& \beta J_{0}^{(\Sigma;l,k)},
%\end{array}
\end{eqnarray}
where we have made use of Eq. (\ref{conn2}) and of the definitions (\ref{THEO1})-(\ref{THEO4}).
It is important to observe, from Eq. (\ref{mapG2}), that, unlike the effective couplings $\beta J^{(\Sigma;l,k)}$,
the couplings $\beta J^{(I;l,k)}$ of the related Ising model are symmetric.

Let us now solve the related Ising model (\ref{HI1}). 
%Notice that, for shortness 
% - from now on - we will omit to write the suffix $I$ on the couplings; 
%the final step being performing the substitutions (\ref{mapG2}).
We have to evaluate the following partition function
\begin{eqnarray}
\label{ZI}
Z_I=\sum_{\{\sigma_i\}}e^{-\beta H_I}.
\end{eqnarray}
Up to terms $\mathop{O}(1)$ $H_I$ can be rewritten as
\begin{eqnarray}
\label{HI2}
H_I &=& H_0\left(\{J_{0}^{(I;l,k)}\};\{h^{(l)}\};\{\sigma_i\}\right)
-\sum_l \frac{J^{(I;l,l)}}{2N\alpha^{(l)}}\left(\sum_{i\in \mathcal{L}_0^{(l)}}\sigma_{i}\right)^2
\nonumber \\
&-& \sum_{l<k} \left[\frac{J^{(I;l,k)}}{2N\alpha^{(l,k)}}\left(\sum_{i\in \mathcal{L}_0^{(l,k)}}\sigma_{i}\right)^2
-\frac{J^{(I;l,k)}}{2N\alpha^{(l,k)}}\left(\sum_{i\in \mathcal{L}_0^{(l)}}\sigma_{i}\right)^2
-\frac{J^{(I;l,k)}}{2N\alpha^{(l,k)}}\left(\sum_{i\in \mathcal{L}_0^{(k)}}\sigma_{i}\right)^2
\right], 
\end{eqnarray}
where we have made use of the definition of $H_0\left(\{J_{0}^{(l,k)}\};\{h^{(l)}\};\{\sigma_i\}\right)$, the
Hamiltonian of the pure model with couplings $\{J_{0}^{(l,k)}\}$ and in the presence of the
external fields $\{h^{(l)}\}$, and we have introduced 
\begin{eqnarray}
\label{Llk}
\mathcal{L}_0^{(l,k)}\defi\mathcal{L}_0^{(l)}\cup \mathcal{L}_0^{(k)}.
\end{eqnarray}

Eq. (\ref{HI2}) can be rewritten also as
\begin{eqnarray}
\label{HI3}
H_I &=& H_0\left(\{J_{0}^{(I;l,k)}\};\{h^{(l)}\};\{\sigma_i\}\right)
-\sum_l \frac{J^{(I;l,l)}}{2N\alpha^{(l)}}\left(\sum_{i\in \mathcal{L}_0^{(l)}}\sigma_{i}\right)^2
\nonumber \\
&-& \sum_{l<k} \frac{J^{(I;l,k)}}{2N\alpha^{(l,k)}}\left(\sum_{i\in \mathcal{L}_0^{(l,k)}}\sigma_{i}\right)^2
+ \sum_l\left[\sum_{k,~k\neq l}
\frac{J^{(I;l,k)}}{2N\alpha^{(l,k)}}\left(\sum_{i\in \mathcal{L}_0^{(l)}}\sigma_{i}\right)^2
\right]. 
\end{eqnarray}
We now proceed analogously to the Ref. \cite{SW} by using the Gaussian transformation to
transform quadratic terms in linear terms coupled to $n+n(n-1)/2$ auxiliary fields 
that we shall indicate with $M^{(l)}$ and $M^{(l,k)}$. 
It is convenient to introduce the following definitions
\begin{eqnarray}
\label{Jtilde}
\frac{\hat{J}^{(I;l,l)}}{\alpha^{(l)}}\defi
\frac{J^{(I;l,l)}}{\alpha^{(l)}}-\sum_{k,k\neq l}\frac{J^{(I;l,k)}}{\alpha^{(l,k)}},
\end{eqnarray}
\begin{eqnarray}
\label{rlk}
r^{(l,k)}\defi
\left\{
\begin{array}{l}
0, \quad \mathrm{if}~ J^{(I;l,k)}\geq 0, \\
1, \quad \mathrm{if}~ J^{(I;l,k)}< 0, 
\end{array}
\right.
\end{eqnarray}
and 
\begin{eqnarray}
\label{rtilde}
\hat{r}^{(l,l)}\defi
\left\{
\begin{array}{l}
0, \quad \mathrm{if}~ \hat{J}^{(I;l,l)}\geq 0, \\
1, \quad \mathrm{if}~ \hat{J}^{(I;l,l)}< 0. 
\end{array}
\right.
\end{eqnarray}
By using these definitions and Eq. (\ref{HI3}), after the Gaussian transformations,
the partition function $Z_I$ reads as
\begin{eqnarray}
\label{ZI1}
Z_I &=&\sum_{\{\sigma_i\}}\int \prod_l dM^{(l)}\prod_{l<k} dM^{(l,k)} 
\exp\left[\beta H_0\left(\{J_{0}^{(I;l,k)}\};\{0\};\{\sigma_i\}\right) \right]
\nonumber \\ 
&\times&\exp\left[-\sum_l \frac{\beta |\hat{J}^{(I;l,l)}|\left(M^{(l)}\right)^2N}{2\alpha^{(l)}}
-\sum_{l<k} \frac{\beta |J^{(I;l,k)}|\left(M^{(l,k)}\right)^2N}{2\alpha^{(l,k)}}\right]
\nonumber \\ 
&\times& \exp\left[\sum_l \left(\frac{\beta\hat{J}^{(I;l,l)}\mathrm{i}^{\hat{r}(l,l)}M^{(l)}}{\alpha^{(l)}}+\beta h^{(l)}\right)
\sum_{i\in\mathcal{L}_0^{(l)}}\sigma_i
+\sum_{l<k} \frac{\beta J^{(I;l,k)}\mathrm{i}^{r(l,k)}M^{(l,k)}}{\alpha^{(l,k)}}
\sum_{i\in\mathcal{L}_0^{(l,k)}}\sigma_i
\right],
\end{eqnarray}
which, by using the definition of $\mathcal{L}_0^{(l,k)}$ and 
$H_0\left(\{J_{0}^{(l,k)}\};\{h^{(l)}\};\{\sigma_i\}\right)$, becomes
\begin{eqnarray}
\label{ZI2}
Z_I &=&\sum_{\{\sigma_i\}}\int \prod_l dM^{(l)}\prod_{l<k} dM^{(l,k)} 
\exp\left[\beta H_0\left(\{J_{0}^{(I;l',k')}\};\{H^{(l')}\};\{\sigma_i\}\right) \right]
\nonumber \\ 
&\times&\exp\left[-\sum_l \frac{\beta |\hat{J}^{(I;l,l)}|\left(M^{(l)}\right)^2N}{2\alpha^{(l)}}
-\sum_{l<k} \frac{\beta |J^{(I;l,k)}|\left(M^{(l,k)}\right)^2N}{2\alpha^{(l,k)}}\right],
\end{eqnarray}
where we have introduced 
\begin{eqnarray}
\label{Heff}
H^{(l)} \defi \frac{\beta\hat{J}^{(I;l,l)}\mathrm{i}^{\hat{r}(l,l)}M^{(l)}}{\alpha^{(l)}}
+\sum_{k,k\neq l} \frac{\beta J^{(I;l,k)}\mathrm{i}^{r(l,k)}M^{(l,k)}}{\alpha^{(l,k)}}
+\beta h^{(l)}.
\end{eqnarray}
For finite $N$ we can exchange the sum over the $\sigma_i$'s with the integral and we get
\begin{eqnarray}
\label{ZI3}
Z_I &=&\int \prod_l dM^{(l)}\prod_{l<k} dM^{(l,k)} e^{-N\mathcal{L}_I(\{M^{(l)}\};\{M^{(l,k)}\})},
\end{eqnarray}
where
\begin{eqnarray}
\label{ZI4}
\mathcal{L}_I\left(\{M^{(l)}\};\{M^{(l,k)}\}\right)&=&
\sum_l \frac{\beta |\hat{J}^{(I;l,l)}|\left(M^{(l)}\right)^2}{2\alpha^{(l)}}
+\sum_{l<k} \frac{\beta |J^{(I;l,k)}|\left(M^{(l,k)}\right)^2}{2\alpha^{(l,k)}}
\nonumber \\
&+&\sum_l\alpha^{(l)}\beta f_0\left(\{\beta J_0^{(I;l',k')}\};\{\beta H^{(l')}\}\right),
\end{eqnarray}
$f_0\left( \{\beta J_0^{(l,k)}\};\{\beta h^{(l)}\}\right)$ being the free energy density of the pure
model with arbitrary couplings $\{J_0^{(l,k)}\}$ and in the presence of arbitrary external
fields $\{h^{(l)}\}$.
By performing the saddle point integration and 
\begin{eqnarray}
\label{ZI5}
\frac{\partial \beta f_0\left(\{\beta J_0^{(I;l',k')}\};\{\beta h^{(l')}\}\right)}{\partial \beta h^{(l)}}=
-m_0^{(l)}\left(\{\beta J_0^{(I;l',k')}\};\{\beta h^{(l')}\}\right) 
\end{eqnarray}
we arrive at the following system of equations for the auxiliary fields
\begin{eqnarray}
\label{ZI6}
\mathrm{i}^{\hat{r}(l,l)}M^{(l)}&=&\alpha^{(l)}m_0^{(l)}\left(\{\beta J_0^{(I;l',k')}\};\{\beta H^{(l')}\}\right), \\
\label{ZI7}
\mathrm{i}^{ r(l,k)}M^{(l,k)}&=&\alpha^{(l)}m_0^{(l)}\left(\{\beta J_0^{(I;l',k')}\};\{\beta H^{(l')}\}\right)
+\alpha^{(k)}m_0^{(k)}\left(\{\beta J_0^{(I;l',k')}\};\{\beta H^{(l')}\}\right),
\end{eqnarray}
where we have used $|J^{(l,k)}|\mathrm{i}^{r^{(l,k)}}/J^{(l,k)}=1/\mathrm{i}^{r^{(l,k)}}$ and 
similarly $|\hat{J}^{(l,l)}|\mathrm{i}^{r^{(l,k)}}/\hat{J}^{(l,l)}=1/\mathrm{i}^{r^{(l,k)}}$.
Eqs. (\ref{ZI6}) and (\ref{ZI7}) lead immediately to identify the auxiliary fields with two indices,
if solution of the saddle point equations, as 
\begin{eqnarray}
\label{ZI8}
\mathrm{i}^{ r(l,k)}M^{(l,k)}=\mathrm{i}^{\hat{r}(l,l)}M^{(l)}+\mathrm{i}^{\hat{r}(k,k)}M^{(k)}.
\end{eqnarray}
If we now use Eqs. (\ref{ZI8}) and the definitions (\ref{Jtilde}) inside Eqs. (\ref{Heff}) 
we see that the $H^{(l)}$'s calculated at the saddle point simplify in
\begin{eqnarray}
\label{Heff1}
%\left. H^{(l)}\right|_{\bm{M}^{\mathrm{sp}}} = \sum_{k} \frac{\beta J^{(I;l,k)}\mathrm{i}^{\hat{r}(k,k)}M^{(k)}}{\alpha^{(l,k)}}
H^{(l)}= \sum_{k} \frac{\beta J^{(I;l,k)}\mathrm{i}^{\hat{r}(k,k)}M^{(k)}}{\alpha^{(l,k)}}+\beta h^{(l)},
\end{eqnarray}
so that the system (\ref{ZI6}) is actually a system of $n$ independent equations in the
$n$ unknowns $M^{(l)}$. 
We can get rid of the imaginary unit by changing the set of variables from $M^{(l)}$
to $\mathrm{i}^{\hat{r}(k,k)}M^{(k)}$. Furthermore, if we divide by $\alpha^{(l)}$, that is if we define
\begin{eqnarray}
\label{ZI9}
m^{(l)}\defi \frac{\mathrm{i}^{\hat{r}(l,l)}M^{(l)}}{\alpha^{(l)}},
\end{eqnarray}
the system (\ref{ZI6}) becomes
\begin{eqnarray}
\label{ZI10}
m^{(l)}=m_0^{(l)}\left(\{\beta J_0^{(I;l',k')}\};\{\beta H^{(l')}\}\right), 
\end{eqnarray}
where the $H^{(l)}$'s, as a function of the fields $m^{(l)}$, have now the form
\begin{eqnarray}
\label{Heff2}
 H^{(l)} = \sum_{k} \frac{\beta J^{(I;l,k)}\alpha^{(l)}m^{(l)}}{\alpha^{(l,k)}}
+\beta h^{(l)}.
\end{eqnarray}
Finally, by performing the mapping transformations (\ref{mapG2})-(\ref{mapG2b}), the system of equations 
(\ref{ZI10})-(\ref{Heff2}) gives the system (\ref{THEO})-(\ref{THEO1}) or, in its most general
form, Eqs. (\ref{THEOG}). Similarly, given a saddle point solution, 
by inserting Eqs. (\ref{ZI9})-(\ref{Heff2}) 
inside Eq. (\ref{ZI4}) we get the free energy density $f_I$ of the related Ising model as
\begin{eqnarray}
\label{Landau}
\beta f_I = L_I\left(\left\{m^{(l')}\right\}\right) +\beta f_0\left(\{\beta J_0^{(I;l',k')}\};\{\beta H^{(l')}\}\right),
\end{eqnarray}
where $L_I$ is defined through $\mathcal{L}_I$ calculated at a given saddle point 
\begin{eqnarray}
\label{Landau1}
L_I\left(\left\{m^{(l')}\right\}\right)=
\mathcal{L}_I\left(\left\{M^{(\mathrm{sp};l)}\left(\left\{m^{(l')}\right\}\right)\right\};
\left\{M^{(\mathrm{sp};l',k')}\left(\left\{m^{(l')}\right\}\right)\right\}\right),
\end{eqnarray}
where $M^{(\mathrm{sp};l)}\left(\left\{m^{(l')}\right\}\right)$ and 
$M^{(\mathrm{sp};l,k)}\left(\left\{m^{(l')}\right\}\right)$ are the given solution of the saddle point equations,
\textit{i.e.}, modulo the definitions (\ref{ZI9}), they satisfy Eqs. (\ref{ZI10})-(\ref{Heff2}).
Finally, by using the mapping transformations (\ref{mapG2})-(\ref{mapG2b}), Eq. (\ref{Landau1})
provides the Landau free energy term (\ref{THEO12}) of the random model, 
or in its more general form Eq. (\ref{THEO12G}).
Of course, as can be checked directly by deriving $\beta f_I$
($\beta f^{(\Sigma)}$, or more simply $L^{(\Sigma)}$), with respect to the external field $\beta h^{(l)}$, 
the saddle point $\{m_I^{(l)}\}$ ($\{m^{(\Sigma;l)}\}$), solution of the system (\ref{ZI10})-(\ref{Heff2})
((\ref{THEOG})), are the local magnetizations $\media{\sigma_i}$ 
($\overline{\media{\sigma_i}^{l_\Sigma}}$), for $i\in\{\mathcal{L}_0^{(l)}\}$, 
of the related Ising model (of the random model).
The non linear system (\ref{THEOG}) may admit many solutions. The unstable solutions which are not
local minima of the functional $\mathcal{L}_I$, do not have any physical meaning and should be discarded by looking
at the Hessian of $\mathcal{L}_I$ equipped with the transformations (\ref{mapG2})-(\ref{mapG2b})
and calculated at the saddle point.
In general we may have more then one stable solution.
The saddle point solution that turns out to be also the absolute minimum of the functional $\mathcal{L}_I$
equipped with the transformations (\ref{mapG2})-(\ref{mapG2b})
corresponds, in the thermodynamic limit, 
to the leading physical solution, the others being metastable states.
Notice that, unlike the case $n=1$, due to the fact that the saddle point solutions
live in a $n$-dimensional section of the original $n+n(n-1)/2$ dimensional space,
$L_I$ turns out to be quite different from $\mathcal{L}_I$ and in particular
the Hessian of $L_I$ has nothing to share with the Hessian of $\mathcal{L}_I$.
Unfortunately the Hessian of $\mathcal{L}_I$, which we recall has to be calculated through the second
derivatives of of $\mathcal{L}_I$ as a function of
a generic point $\{M^{(l)}\},\{M^{(l,k)}\}$ of the $n+n(n-1)/2$ dimensional space,
has a quite complicated form that does not seem to simplify even if calculated at the saddle point.
Note however that, given all the solutions of the self-consistent system
(\ref{THEO})-(\ref{THEO1}) or Eqs. (\ref{THEOG}), for any value of $\beta$, we can always understand which one
is the leading solution: the leading (and of course stable) solution 
is the one that minimizes $L^{(\Sigma)}$. 
\end{widetext} 

%%%%%%%%%%%%%%%%%%%%%%%%

\section{Conclusions}
Motivated by the general issue ``how two given communities influence each other'' discussed in the introduction,
or in other words, ``what are the laws regulating the meta-network'', 
we have formulated the problem through the analysis of a 
generic disordered Ising model built up over a small-world of communities (the meta-network), 
where short-range couplings as well as long-range couplings are completely arbitrary and the graph
disorder is Poissonian.  
By generalizing the method we presented in the Ref. \cite{SW}, we have then solved this random model
in terms of the pure one, where no disorder is present. The resulting self-consistent equations (\ref{THEOG}) 
are a sort of effective TAP equations in which each community contributes as a meta-spin,
as if they were microscopic spins immersed in a ferro or glassy material.    

The consequences of such a general result are then analyzed both at finite and zero temperature 
(in the latter case only for the unfrustrated case). When the number of communities $n$ is not large,
besides ferromagnetism, relative antiferromagnetism among communities may arise if some of the 
long-range couplings $J$'s have negative averages. However, if the number of communities is large, $n\gg 1$,
the TAP-like structure of the equations leads to many metastable states, whose number, in the case in
which the $J$'s have negative averages, may grow exponentially fast with $n$, and a glassy scenario
among communities takes place. In the latter case, 
the system turns out to be highly sensitive to small variations of the set of the many free parameters of the model,
such as the relative sizes of the communities, $\alpha^{(l)}$, the short-range or the long range couplings,
the averages of the added connectivities $c^{(l,k)}$, etc...
In other words, the free energy landscape changes fast by changing the free parameters, so
that many first-order phase transitions are expected when we vary these parameters.
In fact, in a tentative in modeling 
societies, between second- and first-order transitions, the latter are expected to be largely prevalent,
%(in the SK model one has only two free parameters)
consistently with the fact that ``unpredictable'' behavior 
of human communities seems to be a largely prevalent rule.
Finally, at zero temperature the general formula for the relative susceptibilities (\ref{THEOsusc})
has provided us the answer to the fundamental issue ``when two given communities do communicate''.
We find that, unlike the pure model, in the random model two communities $l$ and $k$ do communicate
as soon as $c^{(l,k)}>0$. However, the evaluation of the corresponding characteristic
time $t^{(l,k)}$ depends crucially and in a non trivial way on the susceptibilities of the pure model
in- and between- the two communities via Eqs. (\ref{THEOsusc01})-(\ref{Percotime2}). In particular,
when the matrix of the added connectivities $c^{(l,k)}$ reaches the percolation threshold 
$c_c^{(l,k)}$ determined by Eq. (\ref{P1}), the communities communicate instantaneously (at $T=0$). 

In conclusion, our analysis at finite or zero temperature shows explicitly that methods aimed to study
the communication properties and, more in general, the relationships among communities, of a given network, cannot rely on a local analysis in which
the adjacency matrix is used only in simple algorithms and/or correlations are not taken into account.
Instead, starting from real-data, it
is possible to define in a not ambiguous way a minimal model, a disordered Ising model, able to take into account all the correlations, 
short- and long-range like, present in the given network, and then to capture, via effective TAP equations, the exact relationships
among the communities.

%%%%%%%%%%%%%%%%%%%%%%%%%%%%%%%%%%%%%%%%%%%%%%%%%%%%%%%%%%%%%%%%%%%%
%%%%%%%%%%%%%%%%%%%%%%%%%%%%%%%%%%%%%%%%%%%%%%%%%%%%%%%%%%%%%%%%%%%%

\begin{acknowledgments}
This work was supported by the FCT (Portugal) grants
SFRH/BPD/24214/2005, Socialnets, grant 217141, and PTDC/FIS/71551/2006.
%pocTI/FAT/46241/2002 and
%pocTI/FAT/46176/2003, and the Dysonet Project.
M. O. thanks L. De Sanctis for useful discussions.%having brought to his attention the models faced in this paper.
%We thank A. L. Ferreira and A. Goltsev for useful discussions.
\end{acknowledgments}

\begin{widetext}
\appendix
\section{Partition function for percolation}
The term $-L_0$, as a function of a generic $c$, 
is nothing else than the logarithm of a degenerate partition function 
of the $l$-th community as follows
\begin{eqnarray}
\label{App12}
L_0^{(l)}\left(c\right)=-\lim_{N\to\infty}
\frac{1}{N\alpha^{(l)}}\log\left[\sum_{\{\sigma_i\},i\in\mathcal{L}_0^{(l)}}e^{c\sum_{i\in\mathcal{L}_0^{(l)}}\sigma_i }
\prod_{(i,j)\in\Gamma_0^{(l)}}\left(1+\sigma_i\sigma_j\right)\right],
\end{eqnarray} 
for the case of $J_0^{(l,k)}=0$ for $l\neq k$, whereas in the general case one has
\begin{eqnarray}
\label{App15}
L_0\left(c_1,\ldots,c_n\right)=-\lim_{N\to\infty}
\frac{1}{N}\log\left[\sum_{\{\sigma_i\},i\in\cup_l\mathcal{L}_0^{(l)}}e^{\sum_l c_l\sum_{i\in\mathcal{L}_0^{(l)}}\sigma_i }
\prod_{(i,j)\in\cup_{l,k}\Gamma_0^{(l,k)}}\left(1+\sigma_i\sigma_j\right)\right].
\end{eqnarray} 

\section{Derivation of Eqs. (\ref{Perco4})-(\ref{Perco5})}  
From the definition of the susceptibility, at any temperature we have  
\begin{eqnarray}
\label{Apperco1}
&&\tilde{\chi}_0^{(l,l)}\left(\beta J_0^{(l)};0\right)=\lim_{N\to\infty}
\sum_{i\in\mathcal{L}_0^{(l)}}\frac{\sum_{j\in\cup_l\mathcal{L}_0^{(l)}}
\left[\mediaT{\sigma_i\sigma_j}-\mediaT{\sigma_i}\mediaT{\sigma_j}\right]}{N\alpha^{(l)}}. 
\end{eqnarray}
Now, we have to recall that in our small-world models one has always $\beta_{c0}<\beta_c^{(\mathrm{F})}$,
where $\beta_c^{(\mathrm{F})}$ and $\beta_{c0}$ are the inverse critical temperatures of the model
with and without the added short-cuts, respectively (the random model and the pure model). 
Therefore if the random model is in the P region, also the pure model will
be in its P region. We can repeat the same identical argument for any parameter entering
in our models.
In particular, given $\beta$, if the given connectivity are below the critical percolation surface, not only the 
random model, but also the pure one will be in their P region, so that in such a region Eq. (\ref{Apperco1}) becomes
\begin{eqnarray}
\label{Apperco2}
\tilde{\chi}_0^{(l,l)}\left(\beta J_0^{(l)};0\right)=\lim_{N\to\infty}
\sum_{i\in\mathcal{L}_0^{(l)}}\frac{\sum_{j\in\cup_l\mathcal{L}_0^{(l)}}\mediaT{\sigma_i\sigma_j}}{N\alpha^{(l)}}. 
\end{eqnarray}
Now, due to the fact that in the limit $\beta\to\infty$ two given spins $\sigma_i$ and $\sigma_j$ 
are either infinitely parallel correlated or completely uncorrelated if they are respectively
connected or not by at least a chain of bonds $J_{0;i,j}$
(supposed here only positive if any), from we Eq. (\ref{Apperco2}) we get
\begin{eqnarray}
\label{Apperco3}
\lim_{\beta\to\infty}\tilde{\chi}_0^{(l,l)}\left(\beta J_0^{(l)};0\right)=\lim_{N\to\infty}
\sum_{i\in\mathcal{L}_0^{(l)}}\frac{\mathcal{N}_i^{(l)}}{N\alpha^{(l)}}, 
\end{eqnarray}
where $\mathcal{N}_i^{(l)}$ is defined as the number of vertices belonging to the $l$-th community (including $i$ itself) 
which are reachable from the site $i\in\mathcal{L}_0^{(l)}$ by at least one path of connected vertices.   
Similarly we arrive at Eq. (\ref{Perco5}).
\end{widetext}

%%%%%%%%%%%%%%%%%%%%%%%%%%%%%%%%%%%%%%%%%%%%%%%%%%%%%%%%%%
%%\begin{thebibliography}{99}

\end{document}